\documentclass[psamsfonts, reqno]{amsart}

\usepackage{amssymb,amsfonts}
\usepackage[all,arc]{xy}
\usepackage{enumerate}
\usepackage{subcaption}
\usepackage{enumitem}
\usepackage{graphicx}
\usepackage{mathrsfs}
\usepackage{graphicx}
\usepackage{caption}
\usepackage{subcaption}
\usepackage{fancyref}
\usepackage{dsfont}
\usepackage{scalerel}
\usepackage[affil-it]{authblk}

\usepackage[titletoc]{appendix}
\usepackage{titlesec}

\titleformat{\chapter}    
       {\normalfont\fontsize{16}{19}\bfseries}{\thechapter}{1em}{}

\titleformat{\section}    
       {\normalfont\fontsize{14}{17}\bfseries\filcenter}{\thesection}{0.5em}{}

\titleformat{\subsection}    
       {\normalfont\fontsize{12}{17}\bfseries}{\thesubsection}{0.5em}{}
       
\titleformat{\subsubsection}
       {\normalfont\fontsize{10}{17}\bfseries}{\thesubsection}{0.5em}{}
       
\titlespacing*{\section}
{0pt}{2\baselineskip}{1\baselineskip}

\titlespacing*{\subsection}
{0pt}{2\baselineskip}{1\baselineskip}

\titlespacing*{\subsubsection}
{0pt}{1\baselineskip}{0.5\baselineskip}

\makeatletter
\def\@maketitle{%
  \newpage
  \null
  \vskip 2em%
  \begin{center}%
  \let \footnote \thanks
    {\fontsize{16}{19}\bfseries \@title \par}%
    \vskip 3.5em%
    {\normalsize
      \lineskip .5em%
      \begin{tabular}[t]{c}%
        \@author
      \end{tabular}\par}%
    \vskip 1em%
    {\normalsize \@date}%
    \vskip 0em%
  \end{center}%
  \par
  \vskip 0.5em}
\makeatother

\newcommand\testname{Abstract}
\makeatletter
\newenvironment{cabstract}{%
    \small
    \begin{center}%
        {\bfseries \testname\vspace{0.5em}\vspace{\z@}}%
    \end{center}%
    \quotation}
    {\endquotation}
\makeatother

\usepackage{hyperref}
\usepackage{xcolor}

\hypersetup{pdftitle={main.pdf},
    colorlinks=false,
    linkbordercolor=red
}

\usepackage{geometry}   
\newgeometry{vmargin={30mm}, hmargin={35mm,35mm}}   
  
\usepackage[style=numeric,backend=biber,doi=false,isbn=false,url=false,eprint=false,backref=true]{biblatex}
\usepackage{csquotes}

\usepackage{comment}

\newcommand{\draftcomment}[3]{{\textcolor{#3}{[#2: \emph{#1}]}}}
\renewcommand{\draftcomment}[3]{}  
\newcommand{\brando}[1]{\draftcomment{#1}{\textsc{comment}}{purple}}

\usepackage{enumitem}
\newlist{Properties}{enumerate}{2}
\setlist[Properties]{label=Property \arabic*.,itemindent=*}
\newcommand{\name}{{Surface Augmented Sampler}}

\newcommand{\notation}{\textit{\textbf{Notation.}}}

\newcommand{\vrev}{v_{\text{rev}}}
\newcommand{\wrev}{w_{\text{rev}}}
\newcommand{\zrev}{z_{\text{rev}}}

\theoremstyle{definition}

\usepackage{pgfplots}
\usepackage{pgfplotstable}
\pgfplotsset{width=7cm,compat=1.8}
\theoremstyle{remark}

\makeatletter
\let\c@equation\c@thm
\makeatother
\numberwithin{equation}{section}

\begin{document}

\setlength{\abovedisplayskip}{10pt}   
\setlength{\belowdisplayskip}{10pt}    

\title[Monte Carlo with Soft Constraints]{Monte Carlo with Soft Constraints: \\ the Surface Augmented Sampler}

\author{BRANDO MAGNANI}
\address{Courant Institute of Mathematical Sciences, New York University, 251 Mercer Street, New York, NY 10012, USA}
\email{im975@cims.nyu.edu}
\maketitle

\begin{cabstract}
We describe an MCMC method for sampling distributions with {\em soft} constraints, which are constraints that are almost but not exactly satisfied. We sample a total distribution that is a convex combination of the target "soft" distribution with the nearby "hard" distribution supported on the constraint surface. Hard distribution moves lead to performance that is uniform in the softness parameter. On/Off moves related to the Holmes-Cerfon Stratification Sampler enable sampling the target soft distribution. Computational experiments verify that performance is uniform in the soft constraints limit.
\end{cabstract}

\section{Introduction}\label{sec:sec1}

Many physical and engineering systems involve motion with \textit{soft} constraints, that is, with constraints that are only approximately satisfied. This type of constraints arise in the systems' low temperature limit, and correspond to a situation in which motion occurs (with high probability) within a small neighborhood of a surface that is given as a set of equality constraints. On the other hand, \textit{hard} constraints are those which are imposed directly on the equations of motion of the system.

When simulating particles interacting with bonds that can break and form, one often treats the bond lengths as fixed (as hard constraints) to speed up sampling from the equilibrium distribution. This ad hoc assumption is oftentimes not physically realistic. For example, in molecular dynamics bonds between particles are never fixed at one length, but rather vibrate around that length (an example of soft constraints). The effects of this vibration are not well studied because the available sampling methods either can't handle soft constraints or, if they can, they require a step size that is much smaller than the scale at which it is possible to observe interesting rearrangement of the system. 

This paper will introduce the \name\ algorithm, a Markov chain Monte Carlo method that efficiently samples distributions which arise from the imposition of soft constraints to a given physical system. We are dealing with the the problem of sampling from distributions that are concentrated around a smooth $d$-dimensional surface embedded in a $d_a$-dimensional ambient space. Suppose we have $m$ smooth functions $q_i :\mathbb{R}^{d_a} \rightarrow{\mathbb{R}}$, $i=1,...,m$. Together, these define a \textit{constraint function} $q:\mathbb{R}^{d_a}\rightarrow\mathbb{R}^m$ and a \textit{hard constraint surface}
\begin{align} \label{eq:1.1}
    S = \big\{ x\in\mathbb{R}^{d_a} \ : \ q(x) = 0 \big\}. 
\end{align}
Throughout the rest of the paper, we will use $\nabla q(x)$ to denote the matrix whose colums are the gradients $\{\nabla q_i(x)\}_{i=1}^m$. This matrix corresponds to the transpose of the Jacobian of the full constraint function $q(x)$. We will assume that $\nabla q(x)$ has full column rank $m$ everywhere in $\mathbb{R}^{d_a}$. The implicit function theorem then implies that the dimension of $S$ is $d=d_a-m$ and that the tangent space $\mathcal{T}_x \equiv \mathcal{T}_x S$ is well-defined for every $x\in S$. In this case, the vectors $\{\nabla q_i(x)\}_{i=1}^m$ form a basis for the normal space $\mathcal{N}_x \equiv \mathcal{T}_x S^\perp$. We will also use $\sigma(dx)$ to denote the $d$-dimensional Hausdorff measure on $S$.

We are interested in sampling from the distribution
\begin{align} \label{eq:1.2}
    \pi_\varepsilon(dx) = \frac{1}{Z_\varepsilon} \exp\Big(-\frac{U(x)}{2\varepsilon^2}\Big)dx, \quad U(x) = |q(x)|^2 = \sum_{i=1}^m q_i(x)^2, 
\end{align}
where $\varepsilon$ is a fixed parameter and $Z_\varepsilon = \int_{\mathbb{R}^{d_a}}\exp(-\frac{U(x)}{2\varepsilon^2})dx $. Note that $\varepsilon$ determines the degree to which the distribution is concentrated near $S$. We want to be able to efficiently sample \eqref{eq:1.2} in the \textit{soft constraints limit}, that is, as the temperature (or equivalently, $\varepsilon$) is taken to be arbitrarily small, in which case the distribution is highly concentrated near $S$. The problem of sampling in the soft constraints limit is generally difficult. A typical sample $X$ from \eqref{eq:1.2} will have $\text{dist}(X,S) = O(\varepsilon)$. Therefore, any standard method would require an $O(\varepsilon)$ step size which, in turn, would result in highly correlated samples. Physically, this means that the step size is much smaller than the scale at which interesting rearrangements of a given system can be observed. Our final goal is to construct a sampler that is \textit{effective} in the sense that it produces chains whose auto-correlation time stays bounded as $\varepsilon \rightarrow 0$.  

We note here that, at low temperatures, $\pi_\varepsilon(dx)$ behaves approximately as a degenerate Gaussian distribution. If we Taylor expand $q(x)$ around a point $x_s \in S$, then for (say) $|x-x_s| \leq \varepsilon $ we have
\begin{align} \label{eq:1.3}
    U(x) \approx (x-x_s)^T \nabla q(x) \nabla q(x)^T (x-x_s). 
\end{align}
The degeneracy comes from the fact the rank of $\nabla q(x) \nabla q(x)^T$ is $m < d_a$. A related result is that in the limit as $\varepsilon\rightarrow 0$ we have
\begin{align} \label{eq:1.4}
     \int_{\mathbb{R}^{d_a}} \varphi(x) \pi_\varepsilon(dx) \longrightarrow \int_{S} \varphi(x) \pi_s(dx),
\end{align}
for any smooth test function $\varphi$ (see appendix \ref{sec:A}), where 
\begin{align} \label{eq:1.5}
    \pi_s(dx) \propto \det(\nabla q(x)^T \nabla q(x))^{-1/2} \sigma(dx). 
\end{align}
The presence of the gradient factor in \eqref{eq:1.5} is what distinguishes hard constraints from soft constraints. While hard constraints are imposed directly on the equations of motion of a given system, soft constraints arise as we take the temperature to be arbitrarily small. The fact that $\pi_s \neq \sigma$ shows that the statistical properties of soft constraints are fundamentally different from those of hard constraints.

The \name\ can be seen as an extension to either the ZHG Surface Sampler \cite{ZHG18} or the Stratification Sampler \cite{H20}. In addition to surface moves, our method sometimes proposes jumps to points that are off the constraint surface $S$. This move is similar to the "gain" move in \cite{H20}, with the difference that our method attempts to drop all the constraints at once, whereas a "gain" move drops only one of the constraints at a time. If the current state of the chain is a point that does not lie on the constraint surface $S$, our method proposes to either take an isotropic Gaussian Metropolis step in $\mathbb{R}^{d_a}$ or to move back to $S$. A move back to $S$ is similar to the "lose" move in \cite{H20}, the difference being that our method adds all the constraints at once, while a "lose" move only adds one constraint at a time. Overall, our sampler randomizes  between four different move types. The idea is that one of them, the Surface Sampler move, will have an $O(1)$ step size as $\varepsilon\rightarrow 0$. This should substantially reduce the auto-correlation within the sample.

In order to implement these ideas, we need to work with an $\textit{augmented}$ distribution that allows us to combine surface moves along $S$ with moves that sample points in $S^c$. We denote the augmented distribution by $\rho$ and define it as
\begin{align} \label{eq:1.6}
    \rho(dx) \propto \pi_\varepsilon(dx) + \pi_s(dx). 
\end{align}
The \name\ will preserve distribution \eqref{eq:1.6}. This  means that, at equilibrium, samples in $S^c$ will be distributed according to $\pi_\varepsilon$, while samples in $S$ will be distributed according to $\pi_s$. We decided to "weight" the hard constraint part of $\rho$ by $\pi_s$ not only because it's the physically correct measure in the soft constraints limit, but also because this will help us achieve $100\%$ acceptance probabilities in case $S$ is a flat manifold (more on this in Section \ref{sec:sec3}). 

\textit{\textbf{Acknowledgments.}} Thank you to my advisor Jonathan Goodman and to Miranda Holmes-Cerfon for the helpful discussions. Also thank you to Georg Stadler for reading earlier drafts of this work.

\section{Algorithm}\label{sec:sec2}

In this section we give an overview of our algorithm, the \name, for generating a Markov Chain $X_1, X_2, ... \in \mathbb{R}^{d_a}$ with stationary distribution $\rho$ as in \eqref{eq:1.6}. We will rewrite distribution \eqref{eq:1.6} in a slightly more abstract form that will (hopefully) simplify the forthcoming description. For this, we need to introduce the following notation.

\vspace{3mm}
\notation\ Let $\Omega_1= S^c = \mathbb{R}^{d_a}\setminus S$, $\Omega_2 = S$, and $\Omega =\Omega_1\cup\Omega_2=\mathbb{R}^{d_a}$. Additionally, let $\mu_1(dx)=dx$, the Lebesgue measure on $\mathbb{R}^{d_a}$, and $\mu_2(dx)=\sigma(dx)$, the $d$-dimensional Hausdorff measure on $S$. Lastly, let $f_1(x) = k_1 \exp(-\frac{U(x)}{2\varepsilon^2})$ and $f_2(x) = k_2 \det(\nabla q(x)^T \nabla q(x))^{-1/2}$, where $k_1, k_2$ are constants that will be determined in Section \ref{sec:sec3}.
\vspace{3mm}

One observation is that the surface set $S$ has Lebesgue measure zero, so we may think of the two measures $\mu_1$ and $\mu_2$ as being mutually singular on the whole space $\Omega$. We can now rewrite \eqref{eq:1.6} as 
\begin{align} \label{eq:2.1}
    \rho(dx)&= \frac{1}{Z} \Big[ f_1(x) \mu_1(dx) +  f_2(x) \mu_2(dx) \Big], 
\end{align}
where $Z = \int_{\Omega_1} f_1(x) \mu_1(dx)+\int_{\Omega_2} f_2(x) \mu_2(dx)$. 

The $k$-th state of the Markov chain generated by the \name\ is represented as a pair $(X_k, i_k)$, where $X_k \in \Omega_{i_k}$ and $i_k\in \{1, 2\}$. Suppose that the $k$-th state is $(X_k, i_k) = (x, i)$. The next point $(X_{k+1}, i_{k+1})$ is generated by first constructing a proposal $(y,j)$ and then accepting or rejecting it by Metropolis-Hastings. 

The proposal is constructed via the following two steps:
\vspace{2mm}
\begin{enumerate}
    \item Choose a new label $j$ with probability $\lambda_{ij}(x)$; \label{it:1}
    \\
    \item Chooose a new point $y \in \Omega_j$ with probability density $h_{ij}(x,y)$ with respect to $\mu_j(dy)$. \label{it:2}
\end{enumerate}
\vspace{2mm}
In step \eqref{it:1}, $\lambda_{ij}(x)$ represents the probability of choosing label $j$ given the current state $(x, i)$. Note that we must have $\lambda_{i1}(x) + \lambda_{i2}(x) = 1$ for any point $x\in \Omega_i$ and label $i\in\{1,2\}$. The function $h_{ij}(x,y)$ in step \eqref{it:2} is the probability density of proposing point $y\in\Omega_j$ given the that the current state is $(x, i)$ and that we chose a new label $j$.

The proposal $(y,j)$ is accepted with probability $a_{ij}(x,y)$ defined by
\begin{align} \label{eq:2.2}
a_{ij}(x,y) = \min\Bigg\{1, \ \frac{f_j(y)\lambda_{ji}(y)h_{ji}(y,x)}{f_i(x)\lambda_{ij}(x)h_{ij}(x,y)}\Bigg\}.
\end{align}
The algorithm generates $U\sim \text{Unif}([0,1])$ and if $U<a_{ij}(x,y)$ then we accept the proposal and set $(X_{k+1}, i_{k+1}) = (y,j)$, otherwise we reject it and set $(X_{k+1}, i_{k+1}) = (x,i)$.

We will show in Section \ref{sec:sec4} that the Metropolis-Hastings acceptance probability leads to a Markov transition kernel that is self-adjoint in $L^2_\rho(\Omega)$, which implies that the chain preserves the target distribution $\rho$. Hence, we expect that for reasonable choices of proposal densities the Markov chain is ergodic. If $X_k \sim \rho_k$, this means that for any initial condition we have $\rho_k\rightarrow \rho$ as $k \rightarrow \infty$.

\textit{\textbf{Important Remarks:}} \brando{This was modified} Some additional considerations are required to implement the \name. The first is that when $i\neq j$ the proposal densities in \eqref{eq:2.2} are taken with respect to different reference measures, therefore one has to account for the Jacobian of the maps between $\Omega_i$ and $\Omega_j$. Additionally, the proposal has to be constructed so that there is a unique map between the current state $(x, i)$ and the proposal $(y, j)$. Last but not least, we want to find proposal densities that produce an average acceptance probability that is neither too high nor too low. This is non-trivial when $i\neq j$, that is, when the sampler proposes to add or drop the constraints. The corresponding moves need to be constructed so that the numerator and the denominator in the Metropolis-Hastings ratio in \eqref{eq:2.2} are asymptotically equivalent in the soft constraints limit ($\varepsilon\downarrow 0$). If not, we would have either $a_{12}$ or $a_{21}$ vanishing uniformly as $\varepsilon \downarrow 0$, which in turn would make the \name\ very inefficient\footnote{Very inefficient because in this case transitioning from the $\pi_s$-regime to the $\pi_\varepsilon$-regime (or viceversa) becomes a \textit{rare event}. If the sampler gets stuck in the $\pi_s$-regime for a long time, then many of the generated samples will be discarded since we are only interested in draws from $\pi_\varepsilon$. On the other hand, if the sampler gets stuck in the $\pi_\varepsilon$-regime, we will retain the samples but these will be highly correlated since most of them will be generated via isotropic Gaussian Metropolis with an $O(\varepsilon)$ step size (see the Soft move in Section \ref{sec:subsec2.2})}. As we will see in Section \ref{sec:sec3}, one way to achieve such asymptotic equivalence is to construct moves that produce a $100\%$ acceptance probability when $S$ is a flat manifold defined by linear constraints (such as a line or a plane). When the surface is not flat, the smoothness of the constraint function $q$ then guarantees that $S$ is locally well-approximated by a plane (its tangent space), so that we can expect $a_{12}(x,y), a_{21}(x,y) \nearrow 1$ as $\varepsilon \downarrow 0$. This will allow us to control the average acceptance probability independently of the size of $\varepsilon$.

\subsection{Label proposal and move type}\label{sec:subsec2.1} We will distinguish between four different move types. The new label proposal $j$ and the current label $i$ will determine the move type as follows (we recall here that $\Omega_1 = S^c$ and $\Omega_2 = S$):
\vspace{2mm}
\begin{itemize}
    \item \textit{Hard move}: We choose $j=i=2$. That is, $x, y \in \Omega_2$. \\
    This means that we take a step along the hard constraint $S$. Clearly, this move keeps all the constraints active.
    \vspace{3mm}
    \item \textit{Off move}: $i=2$ and choose $j=1$. That is, $x \in \Omega_2$ and $y \in \Omega_1$.\\
    This means that we have current state $x\in S$ and propose a point $y\in S^c$. This corresponds to a jump "off" the hard constraint, which is equivalent to "dropping" all the constraints at once. 
    \vspace{3mm}
    \item \textit{On move}: $i=1$ and choose $j=2$. That is, $x \in \Omega_1$ and $y \in \Omega_2$.\\
    This means that we have current state $x\in S$ and propose a point $y\in S^c$. This corresponds to a jump back "on" the hard constraint, which is equivalent to "adding" all the constraints at once. 
    \vspace{3mm}
    \item \textit{Soft move}: We choose $j=i=1$. That is, $x, y \in \Omega_1$. \\
    This amounts to a step in  $S^c$.
\end{itemize}
\vspace{2mm}

A practical way to propose a new label is the following. We first decide the overall probabilities $\lambda_{11}, \lambda_{12}, \lambda_{21}, \lambda_{22} \geq 0$ such that $\lambda_{11} + \lambda_{12} = 1$ and $\lambda_{21} + \lambda_{22} = 1$. Suppose that the current state is $(x,i)$. At this point, the algorithm draws $U\sim \text{Unif}([0,1])$ and 
\vspace{2mm}
\begin{itemize}
    \item if $i=1$ and $U<\lambda_{11}$, then we choose a Soft move. Else, if $U>\lambda_{11}$ we choose an On move.
    \vspace{3mm}
    \item if $i=2$ and $U<\lambda_{22}$, then we choose a Hard move. Else, if $U>\lambda_{22}$ we choose an Off move.
\end{itemize}
\vspace{2mm}
Note that this is equivalent to setting $\lambda_{ij}(x) = \lambda_{ij} \mathds{1}_{\Omega_i}(x)$ for all pairs $i,j \in \{1, 2\}$. This implies that $\sum_{i,j} \lambda_{ij}(x) = 1$ for any $x\in \Omega$.

\subsection{Point proposal}\label{sec:subsec2.2} Once we have chosen a label $j$, we must propose a point $y\in \Omega_j$. We now present a proposal mechanism for each of the four move types.

As we will see, three of the four moves depend on taking steps in the tangent space of the hard constraint surface $S$. We use $\mathcal{T}_x$ to denote the tangent space of $S$ at a point $x\in S$. Recall that $\mathcal{T}_x$ is the $d$-dimensional space of vectors $v\in \mathbb{R}^{d_a}$ such that $\nabla q(x)^T v = 0$. This represents the linear approximation to $S$ near $x$. The normal space of $S$ at point $x$ is denoted by $\mathcal{N}_x = \mathcal{T}_x^\perp$ and it is defined as the $m$-dimensional space of vectors $v\in \mathbb{R}^{d_a}$ which are orthogonal to $\mathcal{T}_x$.

We will use $T_x$ to denote an orthonormal matrix such that $\text{Col}(T_x) = \mathcal{T}_x$. The matrix $T_x$ can be computed by taking the last $d$ columns in the QR decomposition of $\nabla q(x)$. 

\subsubsection*{Hard move: $x, \ y \in \Omega_2.$}\label{sec:subsubsec2.2.1} 

To generate a point $y\in S$ we use the ZHG Surface Sampler \cite{ZHG18}. We start by taking a tangential step $x \mapsto x + v$ with $v \in \mathcal{T}_x$. We generate $v$ by sampling a degenerate isotropic Gaussian defined on the linear subspace $\mathcal{T}_x$ with variance $\sigma_\text{hrd}^2$. One way to sample $v$ is to take $v = T_x R$, where $R\in \mathbb{R}^d$ is a random vector with i.i.d. entries $R_i \sim N(0, \sigma_\text{hrd}^2)$, for $i=1,...,d$. The density for $v$ is given by
\begin{align} \label{eq:2.3}
    p_\text{hrd}(v) = \frac{1}{(2\pi)^{d/2} \sigma_\text{hrd}^{d}} e^{- \frac{|v|^2}{2\sigma_\text{hrd}^{2}} }.
\end{align}
Since the tangent step usually brings us off the surface, we need to project back onto $S$. Given $x$ and $v$, the projection step finds a $w\in \mathcal{N}_x$ such that $x + v + w \in S$. This is done by first finding a vector $a \in \mathbb{R}^m$ such that
\begin{align}\label{eq:2.4}
    q\big(x+v+ \nabla q(x) a\big) = 0,
\end{align}
and then setting $w = \nabla q(x) a$. The point proposal will be $y = x + v + w$. 

In our code, we numerically solve equation \eqref{eq:2.3} by standard Newton's method. There are two cases when we must reject the proposal $(y,2)$ so that detailed balance is preserved. First, if Newton's method for \eqref{eq:2.3} does not converge, we reject $(y,2)$ immediately and set $(X_{k+1}, i_{k+1}) = (x,2)$. Second, we must check whether the reverse move is possible, that is, whether we can propose point $x$ starting from $y$. The reverse move is constructed as $x = y + v' + w'$ where $v' \in \mathcal{T}_y$ and $w' \in \mathcal{N}_y$. Since we already know both $x$ and $y$, we can always find $v'$ and $w'$ by projecting $x-y$ onto $\mathcal{T}_y$ and $\mathcal{N}_y = \mathcal{T}_y^\perp$. Then, we must check whether the Newton solver would find $x$ starting from $y+v'$. If not, we reject $(y,2)$ and set $(X_{k+1}, i_{k+1}) = (x,2)$.

The density for the point proposal $y$ is given by $h_{22}(x,y) = p_\text{hrd}(x,v) J(x,y)$, where $J(x,y)$ is the inverse of the determinant of the Jacobian of the projection map $v \rightarrow y$. The Jacobian factor accounts for how the area of a small patch near $x$ is distorted upon mapping it to $y$. One can linearize the projection map to find that $J(x,y) = \det(T_x^T T_y)$ (see Section \ref{sec:subsec2.4} for details on how to compute the Jacobian). The point proposal density for the Hard move is thus given by
\begin{align} 
\begin{split} \label{eq:2.6}
       h_{22}(x,y) &= p_\text{hrd}(v) \det(T_x^T T_y) \\ 
       \\
       & = \frac{1}{(2\pi)^{d/2} \sigma_\text{hrd}^{d}} e^{- \frac{|v|^2}{2\sigma_\text{hrd}^{2}} } \det(T_x^T T_y).
\end{split}
\end{align}
Note that in practice we do not need to compute $\det(T_x^T T_y)$. This is because the Jacobian for the reverse move is $J(y,x) = \det(T_y^T T_x) = \det(T_x^T T_y) = J(x,y)$, so that the Jacobian factors simplify in the Metropolis-Hastings ratio.

\subsubsection*{Off move: $x\in \Omega_2, \ y\in\Omega_1.$}

Given $x\in S$, an Off move generates a point proposal $y\in S^c$. One possibility would be to take an isotropic Gaussian step in $\mathbb{R}^{d_a}$. As we will see  below, this becomes unfeasible in practice because it is impossible to make such proposal symmetric with the On move. By contrast, we take a step $v$ that has different sizes in the directions normal to and tangential to the boundary of $S$. The point proposal will then be $y = x + v$. For the step in the normal direction, we first need the transpose of the pseudo-inverse of $\nabla q(x)$, which we denote by $N_x$. If $\nabla q(x) = U \Sigma V^T$  is the singular value decomposition for $\nabla q(x)$, then $N_x = U \Sigma^{-1} V^T$, where $\Sigma^{-1}$ is obtained by taking the reciprocals of the non-zero diagonal elements in $\Sigma$. We construct $v$ by combining a random step in the directions spanned by $N_x$ and a random step in the directions spanned by $T_x$, so that
\begin{align}
\begin{split} \label{eq:2.7}
    v = N_x R_n + T_x R_t,
\end{split}
\end{align}
where $R_n\in \mathbb{R}^m$ is a random vector with i.i.d. entries $R_{n,i} \sim N(0, \sigma_\text{prp}^2)$ and $R_t\in \mathbb{R}^d$ is a random vector with i.i.d. entries $R_{t,i} \sim N(0, \sigma_\text{tan}^2)$. 

Note that $\text{col}(N_x) = \mathcal{N}_x$, however $N_x$ does not have orthonormal columns. As we will see in Section \ref{sec:sec3}, this particular choice of basis matrix for the step in the normal direction plays a key role in determining a $100\%$ acceptance probability for the Off move in case $S$ is a flat manifold. 
The point proposal density for the Off move is denoted by
\begin{align*}
    h_{21}(x,y) = p_{\text{off}}^{\perp}(v_n) p_{\text{off}}^{\|}(v_t),
\end{align*}
where $p_{\text{off}}^{\perp}(v_n)$ is the density for the normal step $V_n = N_x R_n$, and $p_{\text{off}}^{\|}(v_t)$ is the density for the tangent step $V_t = T_x R_t$. 

Observe that $V_n$ is a degenerate Gaussian vector with mean zero and covariance matrix 
\begin{equation}
    \begin{split}\label{eq:2.8}
        C &= \sigma_\text{prp}^2 N_x N_x^T  \\
        &= \sigma_\text{prp}^2 U \Sigma^{-1} \Sigma^{-T}U^T 
    \end{split}
\end{equation}
(this is degenerate since the rank$(N_x N_x^T)$ = rank$(\nabla q(x) \nabla q(x)^T) = m < d$). Therefore, $V_n = N_x R_n$ has a density
\begin{align} \label{eq:2.9}
    p_{\text{off}}^{\perp}(v_n) = \frac{1}{\sqrt{(2\pi)^m \det^*(C)}}e^{-\frac{1}{2}v_n^T C^+ v_n}
\end{align}
where $C^+$ denotes the pseudo-inverse of covariance matrix $C$ and $\det^*(C)$ denotes its pseudo-determinant (the product of its non-zero singular values). We will verify in Section \ref{sec:subsec2.5} that \eqref{eq:2.9} is equivalent to
\begin{align} \label{eq:2.15}
    p_{\text{off}}^{\perp}(v_n) = \frac{\det(\nabla q(x)^T\nabla q(x)^T)^{1/2}}{(2\pi)^{m/2} \sigma_\text{prp}^{m}} \exp\Big(-v_n^T \frac{\nabla q(x) \nabla q(x)^T}{2\sigma_\text{prp}^2} v_n \Big). 
\end{align}
We will also verify in Section \ref{sec:subsec2.5} that \eqref{eq:2.15} can be written in a more compact form as
\begin{align} \label{eq:2.17}
    p_{\text{off}}^{\perp}(r_n) = \frac{\Pi_{i=1}^m \sigma_i}{(2\pi)^{m/2} \sigma_\text{prp}^{m}} \exp\Big(-\frac{|r_n|^2}{2\sigma_\text{prp}^2} \Big),
\end{align} 
where $v_n = N_x r_n$. By similar arguments, we can write the density for $V_t = T_x R_t$, the step in the tangent direction, as 
\begin{align} \label{eq:2.18a}
    p_{\text{off}}^{\|}(v_t) = \frac{1}{(2\pi)^{d/2} \sigma_\text{tan}^{d}} \exp\Big(-\frac{|v_t|^2}{2\sigma_\text{tan}^2} \Big),
\end{align}
or equivalently
\begin{align} \label{eq:2.18}
    p_{\text{off}}^{\|}(r_t) = \frac{1}{(2\pi)^{d/2} \sigma_\text{tan}^{d}} \exp\Big(-\frac{|r_t|^2}{2\sigma_\text{tan}^2} \Big).
\end{align}
Then, the overall point proposal density for an Off move is given by 
\begin{align} \label{eq:2.19}
\begin{split}
    h_{21}(x,y) &= p_{\text{off}}^{\perp}(r_n) p_{\text{off}}^{\|}(r_t) \\
    \\
    &= \frac{\Pi_{i=1}^m \sigma_i}{(2\pi)^{(m+d)/2} \sigma_\text{prp}^{m}\sigma_\text{tan}^{d}} \exp\Big(-\frac{|r_n|^2}{2\sigma_\text{prp}^2}-\frac{|r_t|^2}{2\sigma_\text{tan}^2} \Big).
\end{split}
\end{align}

Before accepting $y$, we must check that the reverse move is possible. That is, we must check whether it is possible to propose $x$ starting from $y$. This requires knowledge of the On move proposal process, therefore we momentarily postpone this discussion to Section \ref{sec:subsec2.3}. In any case, if the reverse check is successful, we set the proposal to $(y,1)$, otherwise we stop and set $(X_{k+1}, i_{k+1}) = (x,2)$.

\subsubsection*{On move: $x\in \Omega_1, \ y\in\Omega_2.$}\label{sec:subsubsec2.2.3} 

Given $x\in S^c$, an Off move generates a point proposal $y\in S$. We start by projecting $x$ onto the hard constraint $S$. The projection step finds a vector $a_1\in \mathbb{R}^m$ such that  
\begin{align}\label{eq:2.20}
    q\big(x + \nabla q(x) a_1 \big) = 0.
\end{align}
We solve equation \eqref{eq:2.20} by standard Newton's method. If the projection fails, we must immediately stop and set $(X_{k+1}, i_{k+1}) = (x,1)$. If the projection is successful, we take the step $x\rightarrow x_s = x + \nabla q(x) a_1 \in S $ and move on to the rest of the proposal generation process. Note that in the soft constraints limit we can assume that  $\text{dist}(x, S) = O(\varepsilon)$, so that the column space of $\nabla q(x)$ will give a good approximation to the normal space of $S$ at $x_s$, $\mathcal{N}_{x_s}$. If we were to stop here and simply propose $x_s$, we would have a transition kernel that involves only delta-function components. This is because, conditional on $x$, $x_s$ has been generated in a completely deterministic way, which makes detailed balance impossible. Hence, we need to introduce some non-deterministic component to the proposal generation process. Given $x_s$, we generate the proposal by taking a surface move to a point $y = x_s + v + w \in S$, where $v\in \mathcal{T}_{x_s}$ and $w\in \mathcal{N}_{x_s}$. As with the hard move, we first take a step $x_s\rightarrow x_s+v$ with $v\in \mathcal{T}_{x_s}$. We generate $v$ by sampling an isotropic Gaussian in $\mathcal{T}_{x_s}$ with variance $\sigma_{\text{on}}^2$. This tangent space move has a density
\begin{align*}
    p_{\text{on}}(v)  = \frac{1}{(2\pi)^{d/2} \sigma_{\text{on}}^{d}} e^{- \frac{|v|^2}{2\sigma_{\text{on}}^{2}} }
\end{align*}
We then project to $x+v$ back onto $S$ by finding a $w\in \mathcal{N}_{x_s}$ such that $x_s + v + w \in S$. This projection step is carried out by using Newton's method to find an $a_2\in \mathbb{R}^m$ such that $q\big(x_s + v + \nabla q(x_s) a_2 \big)= 0$, and then setting $w =\nabla q(x_s)a_2 \in \mathcal{N}_{x_s}$. Note that here, as for the Hard move, we need to reject the proposal in case the Newton projection step is unsuccessful. The only difference with the Hard move is that in case of an On move there is no need to do the reverse check. In fact, given $x$ and $y$, an Off move from $y$ to $x$ is always possible. We can construct the reverse Off move as $x = y + V'_n + V'_t$ (where $V'_n \in \mathcal{N}_y$ and $V'_t \in \mathcal{T}_y$) by projecting $x-y$ onto $\mathcal{T}_y$ and $\mathcal{N}_y = \mathcal{T}_y^\perp $.

The point proposal density for the On move is given by
\begin{align} 
\begin{split} \label{eq:2.21}
       h_{12}(x,y) &= p_{\text{on}}(v) J(x_s, y) \\ 
       \\
       & = \frac{1}{(2\pi)^{d/2} \sigma_{\text{on}}^{d}} e^{- \frac{|v|^2}{2\sigma_{\text{on}}^{2}} } \det(T_{x_s}^T T_y),
\end{split}
\end{align}
where, as before, $J(x_s, y) = \det(T_{x_s}^T T_y)$ is the Jacobian factor that accounts for the area distortion induced by the projection map $v \rightarrow y$ (see Section \ref{sec:subsec2.4} for details on how to compute the Jacobian).

\subsubsection*{Soft move: $x, \ y\in \Omega_1.$}\label{sec:subsubsec2.2.4}

Given a point $x\in S^c$, we propose $y\in S^c$. For this, we sample an isotropic Gaussian in $\mathbb{R}^{d_a}$ with mean zero and variance $\sigma_\text{sft}^2$. We denote this sample by $v$ and set $y=x+v$. The point proposal density for the Soft move is just
\begin{align} \label{eq:2.22}
    h_{11}(x,y) = \frac{1}{(2\pi)^{d_a/2} \sigma_\text{sft}^{d_a}} e^{-\frac{|v|^2}{2\sigma_\text{sft}^2}}.
\end{align}
Note that the Soft move is just a special case of a Hard move in $\mathbb{R}^{d_a}$.

\subsection{Reverse check for the Off move}\label{sec:subsec2.3} Here we discuss the reverse check process for the Off move. Given $x\in S$, assume that the Off proposal has generated a point $y = x + V_n + V_t \in S^c$, where $V_n \in \mathcal{N}_x$ and $V_t \in \mathcal{T}_x$. We must check whether the On proposal can generate $x$, starting from $y$. The first step is to project $y$ back onto $S$. The reverse projection step attempts to find a vector $a\in \mathbb{R}^m$ such that $ q\big(y + \nabla q(y) a \big) = 0$. If Newton's method fails to converge, we must stop immediately and reject the Off move point proposal $y$. On the other hand, if Newton's method converges to some $y_s = x + \nabla q(y) a \in S$, we need to check whether we can get from $y_s$ to $x$ with a reverse surface move. The reverse surface move is constructed as $x = y_s + v' + w'$ where $v' \in \mathcal{T}_{y_s}$ and $w' \in \mathcal{N}_{y_s}$. Since we already know both $x$ and $y$, we can always find $v'$ and $w'$ by projecting $x-y_s$ onto $\mathcal{T}_{y_s}$ and $\mathcal{N}_{y_s} = \mathcal{T}_{y_s}^\perp$. Thus, we only must check whether the Newton solver would find $x$ starting from $y_s+v'$. If not, we reject $(y,1)$ and set $(X_{k+1}, i_{k+1}) = (x,2)$.

\subsection{Density for the Off move}\label{sec:subsec2.5} In this section we will verify that \eqref{eq:2.9}, \eqref{eq:2.15} and \eqref{eq:2.17} are all equivalent. This will follow from the following four facts. First,
\begin{align} 
    \begin{split}  \label{eq:a}
        C^+ &= \frac{1}{\sigma_\text{prp}^2} \nabla q(x) \nabla q(x)^T,
    \end{split}
\end{align}
second,
\begin{align}   \label{eq:b}
    \det(\nabla q(x)^T\nabla q(x)^T)^{1/2} = \Pi_{i=1}^m \sigma_i,
\end{align}
where the $\sigma_i$'s are the singular values in $\Sigma$. Third,
\begin{align} \label{eq:c}
    \text{det}^*(C) = \frac{\sigma_\text{prp}^{2m}}{\det(\nabla q(x)^T \nabla q(x))},
\end{align}
and fourth,
\begin{align} \label{eq:d}
    |\nabla q(x)^T v_n| = |r_n|,
\end{align}
where $v_n = N_x r_n$. Recall that $\nabla q(x) = U \Sigma V^T$, $N_x = U \Sigma^{-1}V^T$ and $C = \sigma_\text{prp}^2 N_x N_x^T$. We can verify \eqref{eq:a} by
\begin{align*} 
    \begin{split}
        C^+ &= \frac{1}{\sigma_\text{prp}^2} U \Sigma \Sigma^{T}U^T \\
        &= \frac{1}{\sigma_\text{prp}^2} \nabla q(x) \nabla q(x)^T.
    \end{split}
\end{align*}
Let the $\sigma_i$'s be the singular values in $\Sigma$. \eqref{eq:b} holds since 
\begin{align*} 
    \det(\nabla q(x)^T \nabla q(x)) = \det(V^T \Sigma^T \Sigma V) = \Pi_{i=1}^m \sigma_i^2,
\end{align*}
and for \eqref{eq:c} we have
\begin{align*}
\begin{split}
    \text{det}^*(C) & = \text{det}^*(\sigma_\text{prp}^2 U \Sigma^{-1} \Sigma^{-T} U^T)  \\ &= \sigma_\text{prp}^{2m}\text{det}^*(\Sigma^{-1} \Sigma^{-T}) = \frac{\sigma_\text{prp}^{2m}}{\Pi_{i=1}^m \sigma_i^2} = \frac{\sigma_\text{prp}^{2m}}{\det(\nabla q(x)^T \nabla q(x))}.
\end{split}
\end{align*}
We will now verify \eqref{eq:d}. We have
\begin{align*}
    -v_n^T &\nabla q(x) \nabla q(x)^T v_n = - r_n^T N_x^T U \Sigma \Sigma^T U^T N_x r_n = r_n^T r_n.
\end{align*}

\subsection{Calculating the Jacobian}\label{sec:subsec2.4} In this section we will show how calculate the Jacobian factor appearing in the point proposal densities for the Hard and On moves. Similar calculations based on linearization of the projection map are presented in \cite{H20} and \cite{ZHG18}. Let $y = x + v + w$ be the surface sampler proposal point, and suppose $v\rightarrow v + \Delta v$ and $w \rightarrow w + \Delta w$, where $\Delta v\in \mathcal{T}_x$, $\Delta w\in \mathcal{N}_x$ and $\Delta y \in \mathcal{T}_y$. Then, there exists vectors $a, b, c$ such that $\Delta v = T_x a$, $\Delta w = N_x b$ and $\Delta y = T_y c$. We have that $\Delta y = \Delta v + \Delta w$, which can be rewritten as $T_y c = T_x a + N_x b$. If we multiply the last equation by $T_x^T$ we get
\begin{align}
    T_x^T T_y c = a,
\end{align}
where we used that $T_x^T T_x = I$ (by orthonormality of $T_x$) and that $T_x^T N_x  = 0$ (by $\mathcal{N}_x = \mathcal{T}_x^\perp$). The Jacobian of the map $v\rightarrow y$ is thus  given by $(T_x^T T_y)^{-1}$. From this, we conclude that $J(x,y) = (\det(T_x^T T_y)^{-1})^{-1} = \det(T_x^T T_y)$.

\section{Choosing the Correct Scaling: the Flat Surface Case}\label{sec:sec3}

\brando{This is new} In Section \ref{sec:subsec2.2} we found the point-proposal densities for each move type. However, we did not provide any details on how to choose the scale parameters $\sigma_\text{prp}, \sigma_\text{tan}, \sigma_\text{on}, \sigma_\text{hrd}, \sigma_\text{sft}$. In this section we will specialize to the setting where $S$ is a flat manifold defined by linear constraints (such as a line or a plane). In this case, we will see that it is possible to choose a scaling for the On and Off moves that produces a $100\%$ acceptance probability. In other words, we will determine $\sigma_\text{prp}, \sigma_\text{tan}$, and $\sigma_\text{on}$ such that the detailed balance condition
\begin{align} \label{eq:db}
    \begin{split}
        f_2(x)&\lambda_{21}(x)h_{21}(x,y) = f_1(y)\lambda_{12}(y)h_{12}(y,x), \quad \forall \ x \in \Omega_2, y\in \Omega_1
    \end{split}
\end{align}
holds. This will imply that $a_{21}(x,y) = a_{12}(y,x) \equiv 1$. When $S$ is not flat, the smoothness of the constraint function $q$ ensures that the surface is locally well-approximated by a plane. Therefore, we can expect that the same choice of $\sigma_\text{prp}, \sigma_\text{tan}$, and $\sigma_\text{on}$ will lead to $a_{12}(x,y), a_{21}(y,x) \nearrow 1$ as $\varepsilon \downarrow 0$.

We mention here that the scale parameters for the Hard and Soft moves ($\sigma_\text{hrd}$ and $\sigma_\text{sft}$) should be $O(1)$ and $O(\varepsilon)$, respectively. The reason for having a Hard move is that its scaling is not affected by the size of $\varepsilon$, that is, we can have $\sigma_\text{hrd} = O(1)$. This move should substantially de-correlate the samples and produce a method that is effective in the sense of Section \ref{sec:sec1}. Nothing much can be done about the Soft move. As we mentioned before, a typical sample $X$ from \eqref{eq:1.2} will have $\text{dist}(X, S) = O(\varepsilon)$. Therefore, the size for the isotropic Gaussian proposal \eqref{eq:2.22} will have to be to be small, that is, we need to have $\sigma_\text{sft} = O(\varepsilon)$.

\subsection{Detailed balance with a flat surface}\label{sec:subsec3.1} \brando{This was modified in a way that should provide more intuition on how to construct the Off move, given the On move} In this section we will specialize to the flat surface case and find a scaling for the On and Off moves for which condition \eqref{eq:db} holds. Suppose we have a linear constraint function and a surface of the form
\begin{align} \label{eq:3.1}
    q(x) = A^T x , \quad S = \big\{ x\in\mathbb{R}^{d_a} \ : \ q(x) = 0 \big\}.
\end{align}
where $A \in \mathbb{R}^{d_a \times m}$ is assumed to have full column rank $m$. Let $A = U \Sigma V^T$ be the singular value decomposition for $\nabla q(x) = A$, and define the matrix $N = U \Sigma^{-1} V^T$. Note that in this case $\mathcal{N}_x = \text{Col}(A) = \text{Col}(N)$ and $\mathcal{T}_x = \text{Null}(A^T) = S$, for any $x\in S$. We can rewrite the distribution in \eqref{eq:1.2} by,
\begin{align}\label{eq:3.2}
\begin{split}
    \pi_\varepsilon(dx) &= \frac{1}{Z_\varepsilon}\exp\Big(-x^T\frac{ A A^T }{2\varepsilon^2}x\Big) dx.
\end{split}
\end{align}
We see that $\pi_\varepsilon$ is just a degenerate (since $A A^T$ has rank $m<d_a$) Gaussian distribution with covariance matrix $\varepsilon^2 N N^T$ and normalizing constant
\begin{align} \label{eq:3.3}
\begin{split}
    Z_\varepsilon &= \sqrt{(2\pi)^m \varepsilon^{2m} \text{det}^*(N N^T)} \\
    & = \sqrt{(2\pi)^m \varepsilon^{2m} \text{det}^*(\Sigma^{-1} \Sigma^{-T})} \\
    & = \sqrt{(2\pi)^m \varepsilon^{2m} \Pi_{i=1}^m \sigma_i^{-2}} \\
    & = \sqrt{\frac{(2\pi)^m \varepsilon^{2m}}{\det(A^T A)}}.
\end{split}
\end{align}
Therefore, we can rewrite \eqref{eq:3.2} as 
\begin{align}\label{eq:3.4}
\begin{split}
    \pi_\varepsilon(dx) &= \frac{\det(A^T A)^{1/2}}{(2\pi)^{m/2} \varepsilon^{m}}\exp\Big(-x^T\frac{ A A^T }{2\varepsilon^2}x\Big) dx.
\end{split}
\end{align}

Now consider an Off move from $x\in \Omega_2=S$ to point $y\in \Omega_1=S^c$. The reverse On move has the form $x = y_s + \vrev + \wrev $, where $y_s \in S$, $\vrev\in \mathcal{T}_{y_s}$ and $\wrev \in \mathcal{N}_{y_s}$. Recall that $y_s$ is determined by solving the linear system $q\big(y + \nabla q(y) a\big) = 0$ for $a$ via Newton's method. Newton's method finds the solution in one step\footnote{The solution is $a = - (A^T A)^{-1} A^T y$.} and returns $y_s = \text{proj}_S (y) = y + \zrev$, where $\zrev = \nabla q(y) a \in \mathcal{N}_{y_s}$. Since $S$ is flat, we know that the tangent step in the surface move, $\vrev$, will not bring us off the surface. We conclude that the surface move projection step is unnecessary and set $\wrev=0$. To sum up, we have $x = y + \zrev + \vrev$ with $\zrev \in \mathcal{N}_{y_s}$ and $\vrev\in \mathcal{T}_{y_s}$. The point proposal density for the reverse On move is given by $h_{12}(y,x) = p_{\text{on}}(v_{\text{rev}})$, where there is no Jacobian factor since $\det(T_{y_s}^T T_x) \equiv 1$\footnote{Since $S$ is flat, we have $T_x  = T$ and by orthonormality $T_x^T T_y = T_{y_s}^T T_x = T^T T =  I$.}. Detailed balance condition \eqref{eq:db} can be written as
\begin{align}
    \begin{split} \label{eq:3.5}
        h_{21}(x,y) &= \frac{\lambda_{12}(y) f_1(y)}{\lambda_{21}(x)f_2(x)} \ h_{12}(y,x) \\
        \\
        &= \frac{\lambda_{12}}{\lambda_{21}} \ \frac{f_{1}(y)}{f_{2}(x)} \ p_{\text{on}}(v_{\text{rev}}) \\
        \\
        &= \frac{\exp\big(-y^T\frac{ A A^T }{2\varepsilon^2}y \big)}{\frac{\lambda_{21}}{\lambda_{12} k_1} f_2(x) } \ \frac{\exp\big(- \frac{|\vrev|^2}{2\sigma_{\text{on}}^{2}} \big) }{(2\pi)^{d/2} \sigma_{\text{on}}^{d}}, 
    \end{split}
\end{align}
where we used that $f_1(y) = k_1 \exp\big(-y^T\frac{ A A^T }{2\varepsilon^2}y \big)$. Note that \eqref{eq:3.5} suggests how to construct the Off move point proposal so that the detailed balance condition \eqref{eq:db} holds. To balance the two Gaussian terms on the right hand side of \eqref{eq:3.5}, the Off move should consist of a combination of $(i)$ a Gaussian step in the normal space with basis $N_x  = N$ and $(ii)$ an isotropic Gaussian step in tangent space (with any choice of orthonormal basis $T_x = T$). This corresponds to the Off proposal mechanism we presented in Section \ref{sec:subsec2.2}, where the proposal is set to $y = x + v$ with $v = v_n + v_t$ and $v_n = N r_n \in \mathcal{N}_x$ and $v_t = T r_t \in \mathcal{T}_x$. We can use \eqref{eq:2.15}, \eqref{eq:2.18a} to write the Off proposal density as
\begin{align} 
    \begin{split} \label{eq:3.6}
        h_{21}(x,y) &= p_{\text{off}}^{\perp}(v_n) p_{\text{off}}^{\|}(v_t) \\
        \\
        &= \frac{\exp\big(-v_n^T \frac{A A^T}{2\sigma_\text{prp}^2} v_n \big)}{\frac{(2\pi)^{m/2} \sigma_\text{prp}^{m}}{\det(A^T A)^{1/2}}} \ \frac{\exp\big(- \frac{|v_t|^2}{2\sigma_{\text{tan}}^{2}} \big) }{(2\pi)^{d/2} \sigma_{\text{tan}}^{d}},
    \end{split}
\end{align}
where we omitted the Jacobian factor since $\det(T_{x}^T T_y) \equiv 1$. If we substitute \eqref{eq:3.6} into \eqref{eq:3.5} we obtain
\begin{align}
    \begin{split} \label{eq:3.7}
\frac{\exp\big(-v_n^T \frac{A A^T}{2\sigma_\text{prp}^2} v_n \big)}{\frac{(2\pi)^{m/2} \sigma_\text{prp}^{m}}{\det(A^T A)^{1/2}}} \ \frac{\exp\big(- \frac{|v_t|^2}{2\sigma_{\text{tan}}^{2}} \big) }{(2\pi)^{d/2} \sigma_{\text{off}}^{d}} 
= 
\frac{\exp\big(-y^T\frac{ A A^T }{2\varepsilon^2}y\big)}{\frac{\lambda_{21} }{\lambda_{12} k_1} f_2(x) } \ \frac{\exp\big(- \frac{|\vrev|^2}{2\sigma_{\text{on}}^{2}} \big) }{(2\pi)^{d/2} \sigma_{\text{on}}^{d}}.
    \end{split}
\end{align}
Now observe that we must have $\vrev = -v_t$, so that if we choose $\sigma_\text{tan} = \sigma_{\text{on}}$, then the Gaussian densities for the step in tangent space simplify in \eqref{eq:3.7}. Moreover, since $y = x + v_n + v_t$ with $x, v_t \in S = \text{Null}(A^T)$, we have $A^T y = A^T v_n$. Therefore, \eqref{eq:3.7} becomes
\begin{align}
    \begin{split} \label{eq:3.8}
\frac{\exp\big(-v_n^T \frac{A A^T}{2\sigma_\text{prp}^2} v_n \big)}{\frac{(2\pi)^{m/2} \sigma_\text{prp}^{m}}{\det(A^T A)^{1/2}}} 
= 
\frac{\exp\big(-v_n^T\frac{ A A^T }{2\varepsilon^2}v_n\big)}{\frac{\lambda_{21} }{\lambda_{12} k_1} f_2(x) }.
    \end{split}
\end{align}
This last equation tells us the scaling for the normal step component of an Off move. Indeed, to simplify the exponential factors in \eqref{eq:3.8} we have to take $\sigma_\text{prp} = \varepsilon$. Equation \eqref{eq:3.8} also suggests what the density on the surface, $f_2$, should be. In fact, we can rewrite \eqref{eq:3.8} as
\begin{align} \label{eq:3.9}
    \begin{split}
        \frac{f_2(x)}{k_1} &= \frac{ \lambda_{12} (2\pi)^{m/2} \varepsilon^{m}}{ \lambda_{21}}\det(A^T A)^{-1/2}.
    \end{split}
\end{align}
Thanks to our choice of surface density $f_2(x) = k_2 \det(\nabla q(x)^T \nabla q(x))^{-1/2} = k_2 \det(A^T A)^{-1/2}$, the gradient factors simplify. Therefore, \eqref{eq:3.9} becomes
\begin{align} \label{eq:3.10}
    \begin{split}
        \frac{k_2}{k_1} = \frac{ \lambda_{12} (2\pi)^{m/2} \varepsilon^{m}}{ \lambda_{21}},
    \end{split}
\end{align}
which can obviously be reduced to an identity since we are free to choose constants $k_1$ and $k_2$. 

To summarize, we showed that when $S$ is a flat surface, if we set $\sigma_\text{prp} = \varepsilon$, $\sigma_\text{tan} = \sigma_{\text{on}}$ then detailed balance condition \eqref{eq:db} holds, provided that our choice of $k_1, k_2$ satisfies \eqref{eq:3.10}.

\section{Experiments}\label{sec:sec6}

In this section we illustrate the \name\ with two examples of hard constraint surface: the intersection of two spheres and the intersection of an ellipsoid with a sphere. 

For the rest of this section, we will use $X=(X^1, X^2, X^3)$ to denote a random variable in $\mathbb{R}^3$ with distribution $\pi_\varepsilon$ as in \eqref{eq:1.2}. We will also use $x = (x_1, x_2, x_3)$ to denote a generic point in $\mathbb{R}^3$.

In what follows we will estimate the auto-covariance function and the auto-correlation time of MCMC samples $X_1, X_2, ..., X_N$ drawn from $\pi_\varepsilon$. Consider a general estimator 
\begin{align}
    \hat{S}_N = \frac{1}{N} \sum_{j=1}^N F(X_j),
\end{align}
for some given function $F(\cdot)$. Assuming that $\{X_j\}_{j=1}^N$ are sampled from \eqref{eq:1.2}, the equilibrium lag $t$ auto-covariance is
\begin{align}
    C_t = \text{Cov}[F(X_j), F(X_{j+t})].
\end{align}
Note that this is an even function of $t$ since $C_t = C_{-t}$. We estimate the auto-covariance function from the MCMC samples as
\begin{align} \label{eq:covhat}
    C_t \approx \frac{1}{N-t}\sum_{j=1}^{N-t} \big(F(X_j) - \hat{S}_N\big)\big(F(X_{j+t}) - \hat{S}_N\big)
\end{align}
The integrated auto-correlation time, denoted by $\tau$, is defined as 
\begin{align}
    \tau = \sum_{t=-\infty}^{\infty} \frac{C_t}{C_0} = 1 + \frac{2}{\text{Var}(F(X_j))} \sum_{t=1}^\infty C_t
\end{align}
We use the self-consistent window approach (described in \cite{Sok}) to estimate $\tau$. The self-consistent window approach estimates $\tau$ by using a number of samples $M \ll N$. This produces an estimator with lower variance at the cost of increasing the bias. The point of having an accurate estimate for the integrated auto-correlation time is that
\begin{align}
\text{Var}(\hat{S}_N) \approx \frac{\tau}{N} C_0,
\end{align}
so that the error in the estimate decreases as $1/\sqrt{N_\text{eff}}$, where $N_{\text{eff}} = N / \tau$ is the effective sample size. The integrated auto-correlation time is a measure of how many MCMC steps one has to take to generate approximately independent samples.

Note that instead of estimating the auto-covariance function from the sum in \eqref{eq:covhat} (this takes $O(N^2)$ operations), it is computationally more efficient to take the fast Fourier transform of the entire sample (this takes $O(N \log N)$ operations). This uses the fact that the auto-correlation of a continuous-time deterministic signal can be written as a convolution, which in Fourier space corresponds to multiplication.

Note that for most of our experiments we generated approximately $5\times 10^6$ samples from the augmented target distribution  \eqref{eq:2.1}. We used parameters $\lambda_{11} = 0.2$ (Soft move), $\lambda_{12} = 0.8$ (On move), $\lambda_{22} = 0.8$ (Hard move), $\lambda_{21} = 0.2$ (Off move) and $\sigma_\text{prp} = \sigma_\text{tan} = \sigma_\text{on} = \varepsilon$,  $\sigma_\text{hrd} = 1$ and $\sigma_\text{sft} = 0.7 \times \varepsilon$. About $20\%$ of the samples were off the hard constraint surface $S$. That is, with these parameters we obtained $N \approx 10^6$ samples from $\pi_\varepsilon$, which we then used for analysis (samples on the surface $S$ were discarded since these are drawn from $\pi_s$).

\subsection*{Two Spheres model} Consider two soft constraints $q_i : \mathbb{R}^{3} \rightarrow \mathbb{R}$ given by
\begin{align*}
    q_i(x) = |x-c_i|^2 - r_i^2, 
\end{align*}
where $c_1, c_2 \in \mathbb{R}^3$ and $r_1, r_2$ are positive real numbers. Geometrically, the hard constraint surface $S = \big\{ x\in\mathbb{R}^{3} \ : \ q(x) = 0 \big\}$ is the intersection of two spheres with centers $c_1$, $c_2$ and radii $r_1$, $r_2$, respectively. In our experiments, we use centers $c_1 = (0, 0, 1)$, $c_2 = (0, -1, 0)$ and radii $r_1 = r_2 = \sqrt{2}$. In this way, the point $s = (1, 0, 0) \in S$. The surface $S$ consists of a circle with center $c = (0, -1/2, 1/2)$ (the midpoint between $c_1$ and $c_2$) and radius $r = |s-c| = \sqrt{3/2}$. The circle lies on the plane perpendicular to the line segment between $c_1$ and $c_2$, that is, it lies on the plane $P = \big\{ x\in\mathbb{R}^{3} \ : \ (x-p)\cdot \vec{n} = 0 \big\}$, where $\vec{n}=(0,1,1)$. 

In general, this problem is too simple since the gradient factor $\det(\nabla q(x)^T \nabla q(x))$ is constant and the Hard move does not have any Metropolis or projection failures. However, this setting allows for a simple correctness check which involves the angle $\theta$ between the projection of $X\sim \pi_\varepsilon$ onto $P$ and any reference vector perpendicular to $\vec{n}$ (say $w = s-c$). Since in this case the gradient factor in $\pi_s$ is constant, the angle $\theta$ is uniformly distributed. 

\begin{figure}[ht] 
    \centering
  \begin{subfigure}[b]{0.49\textwidth}
    \includegraphics[width=\textwidth]{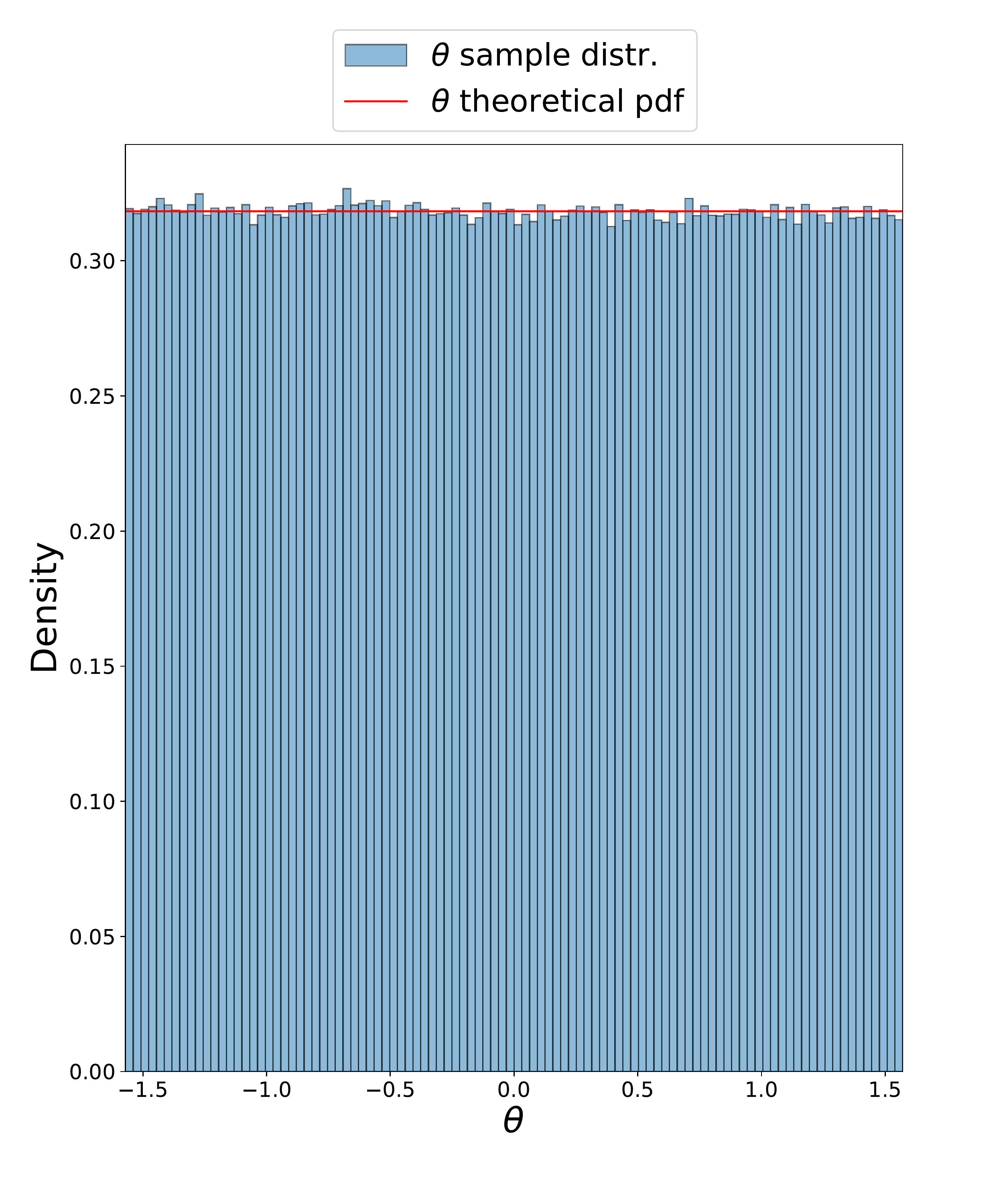}
    \caption{Two-Sphere Model}
  \end{subfigure}
  \begin{subfigure}[b]{0.49\textwidth}
    \includegraphics[width=\textwidth]{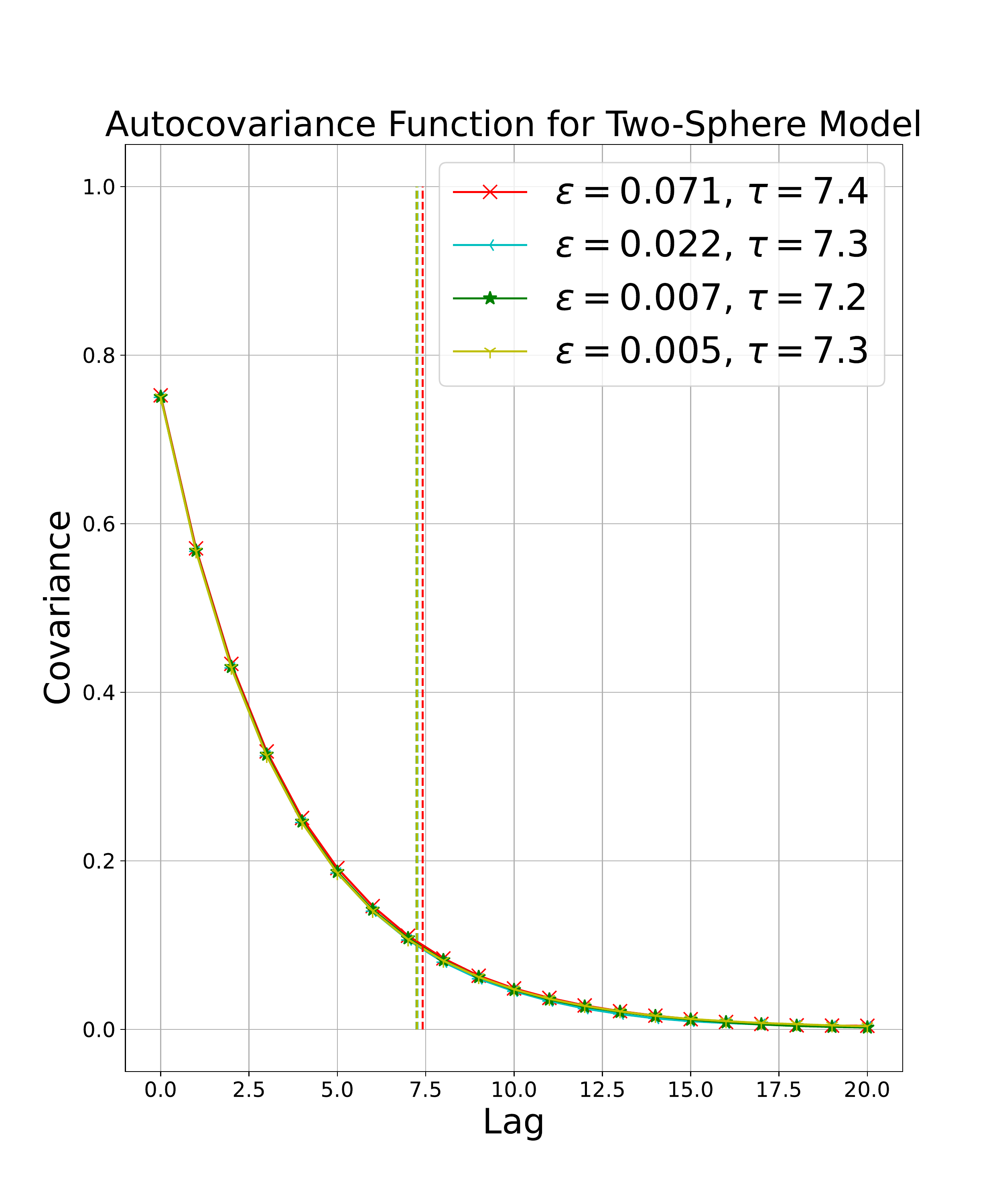}
    \caption{Two-Sphere Model}
  \end{subfigure}
\caption{(A): Sample distribution and theoretical density for $\theta$ with $\varepsilon=0.022$. (B): Sample auto-covariance function and auto-correlation time with $F(x) = x_1$ for decreasing values of $\varepsilon$. }\label{fig:1}
\end{figure}

We performed the $\theta$-check with $\varepsilon=0.022$ (inverse temperature $\beta=1/2\varepsilon^2=10^3$). We computed the angle between the projection of each sample onto $P$ and the reference vector $w$. Figure \ref{fig:1} displays an histogram for the sampled angles. The empirical distribution of the angle $\theta$ looks correct within statistical sampling error. 

We also computed the sample auto-covariance function and auto-correlation time for decreasing values of $\varepsilon$. In particular, we took inverse temperatures $\beta = 1/2\varepsilon^2 = 10^2, 10^3, 10^4, 2\times10^4$ and used $F(x) = x_1$, the projection map onto the first coordinate. Figure \ref{fig:1} contains plots of the auto-covariance structure and lists the integrated auto-correlation time for various values of $\varepsilon$. We see that the auto-correlation time stays approximately constant as $\varepsilon$ decreases.

\begin{figure}[ht] 
    \centering
  \begin{subfigure}[b]{0.4\textwidth}
    \includegraphics[width=\textwidth]{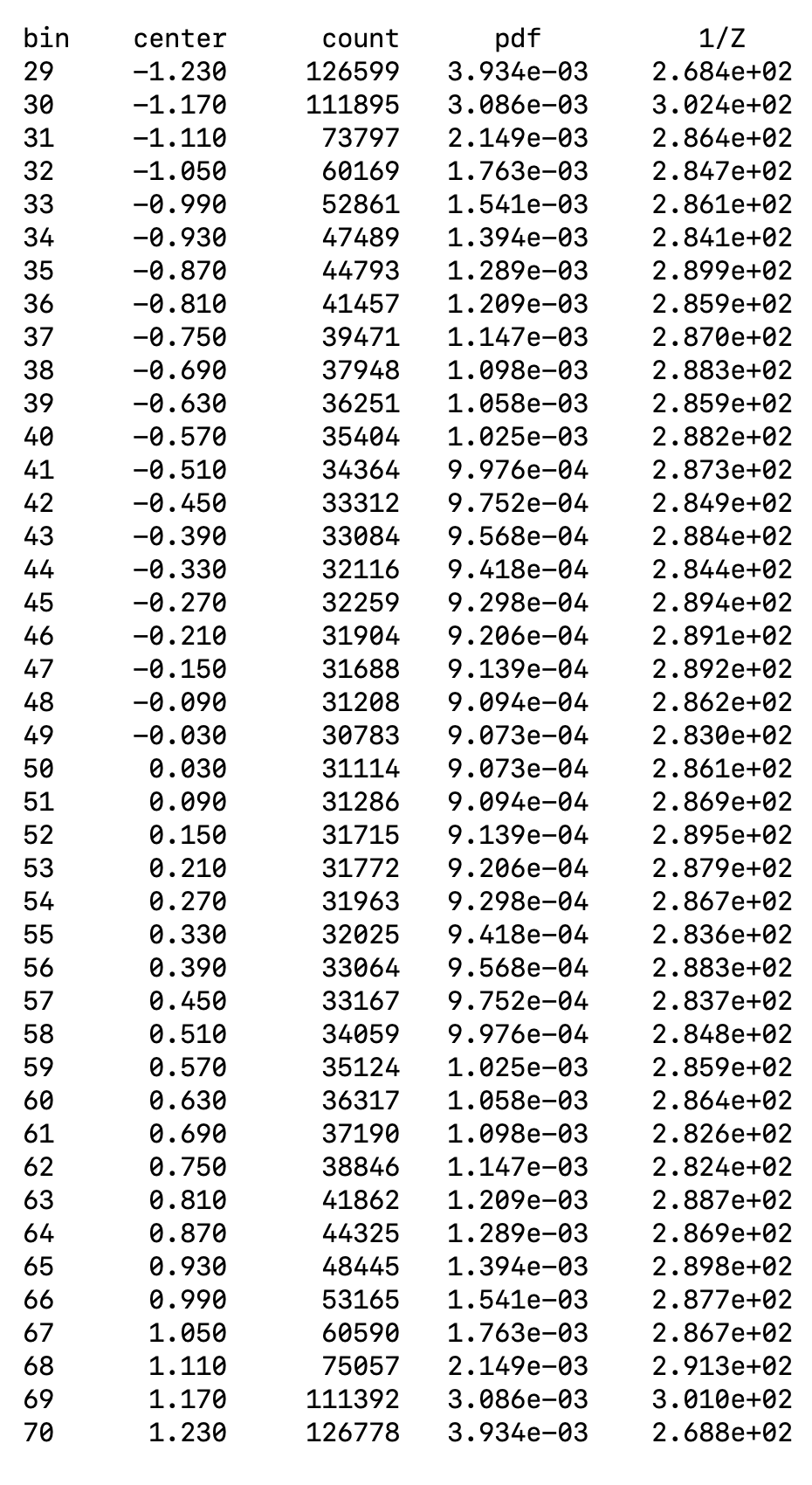}
    \caption{Two-Sphere Model}
  \end{subfigure}
  \hspace{1 cm}
  \begin{subfigure}[b]{0.4\textwidth}
    \includegraphics[width=\textwidth]{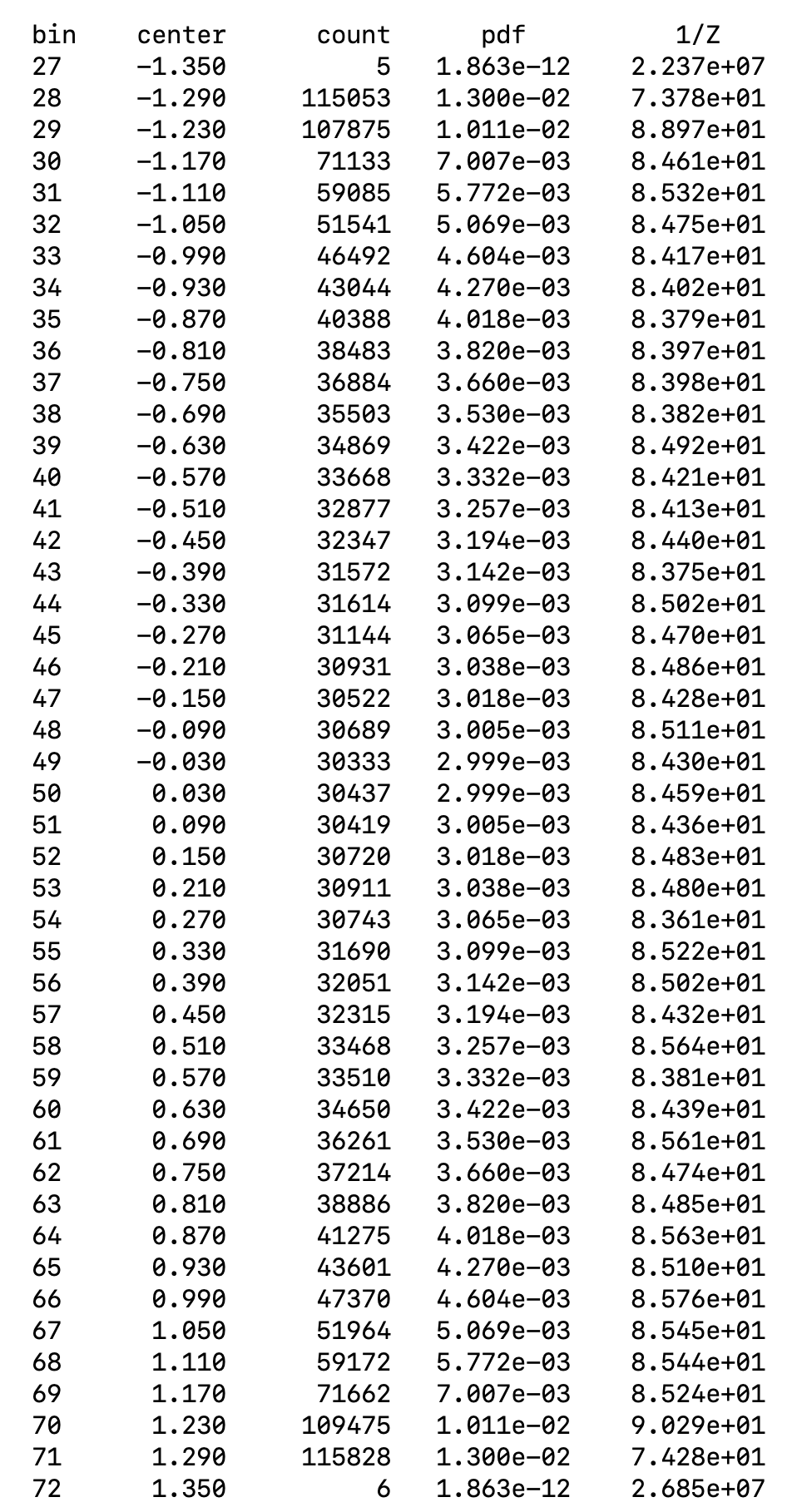}
    \caption{Ellipsoid-Sphere Model}
  \end{subfigure}
\caption{Tables (A) and (B) display bin index ("bin"), center of the bin ("center"), number of bin hits ("count"), exact marginal density at bin center point ("pdf"), and the value of the random variable $\hat{R}^i_N$ ("1/Z"). For both tables, we generated $N \approx 2\times10^6$ samples from $\pi_\varepsilon$ with $\varepsilon=0.022$ (i.e., inverse temperature $\beta=1/2\varepsilon^2 = 1000$).}\label{fig:2}
\end{figure}

As another check, we also estimate the distribution of $X^1$, the first coordinate of $X \sim \pi_\varepsilon$. To do this, we divide the $x_1$-axis into $L$ bins. Note that we can compute the un-normalized density of $X^1$,
\begin{align*}
    p_{X^1}(x_1) = \iint p_X(x_1, x_2, x_3) dx_2 dx_3,
\end{align*}
via numerical quadrature\footnote{We use the midpoint rule.}. We use estimators
\begin{align} \label{eq:ratio}
    \hat{R}_N^i = \frac{\frac{1}{N}\sum_{j=1}^N \mathds{1}_{B_i}(X^1_j)}{p_{X^1}(b_i) |B_i|}, \quad \quad i=1, ..., L
\end{align}
where $B_i$ is the $i$-th bin with center $b_i$ and $X^1_j$ is the first coordinate of the $j$-th sample from the MCMC sampler. If the sampler is correct, we should find that the random variables $\hat{R}_N^i$ are roughly constant over the bin index $i$ since 
\begin{align*}
    \begin{split} 
        \hat{R}_N^i &= \frac{\frac{1}{N}\sum_{j=1}^N \mathds{1}_{B_i}(X^1_j)}{p_{X^1}(b_i) |B_i|} \\
        \\
        &\approx \frac{\text{Pr}\big( X^1 \in B_i \big)}{\int_{B_i} p_{X^1}(x_1) dx_1} = \frac{1}{Z},
    \end{split}
\end{align*}
where $Z = \int p_{X^1}(x_1) dx_1$. This is what we see in Table (A) of Figure \ref{fig:2}.

\subsection*{Ellipsoid-Sphere model} Consider two soft constraints $q_i : \mathbb{R}^{3} \rightarrow \mathbb{R}$ given by
\begin{align*}
    \begin{split}
        q_1(x) &= |x-c_1|^2 - r_1^2, \\
        \\
        q_2(x) &= \frac{(x_1 - c_{2,1})^2}{s_1^2} + \frac{(x_2 - c_{2,2})^2}{s_2^2} + \frac{(x_3 - c_{2,3})^2}{s_3^2} - 1,
    \end{split}
\end{align*}
where $c_1, c_2 \in \mathbb{R}^3$ and $r_1$, $s_1, s_2, s_3$ are positive real numbers. Geometrically, the hard constraint surface $S = \big\{ x\in\mathbb{R}^{3} \ : \ q(x) = 0 \big\}$ is the intersection of a sphere centered at $c_1$ with radius $r_1$ and an ellipsoid centered at $c_2$ (where $s_1, s_2, s_3$ are half the length of its principal axes). In our experiments, we use centers $c_1 = (0, 0, 1)$, $c_2 = (0, -1, 0)$, radius $r_1 = \sqrt{2}$ and $s_1 = 2, s_2 = 3, s_3 = 5$. 

This problem is more complex than the two-spheres model discussed above because  in this case $\det(\nabla q(x)^T \nabla q(x))$ is non-constant. This means that the Hard move will have both Metropolis and reverse check failures. 

We estimated the $X^1$ marginal density in the same way as for the two-sphere model, see Table (B) in Figure \ref{fig:2}.

\begin{figure}[ht] 
    \centering
  \begin{subfigure}[b]{0.49\textwidth}
    \includegraphics[width=\textwidth]{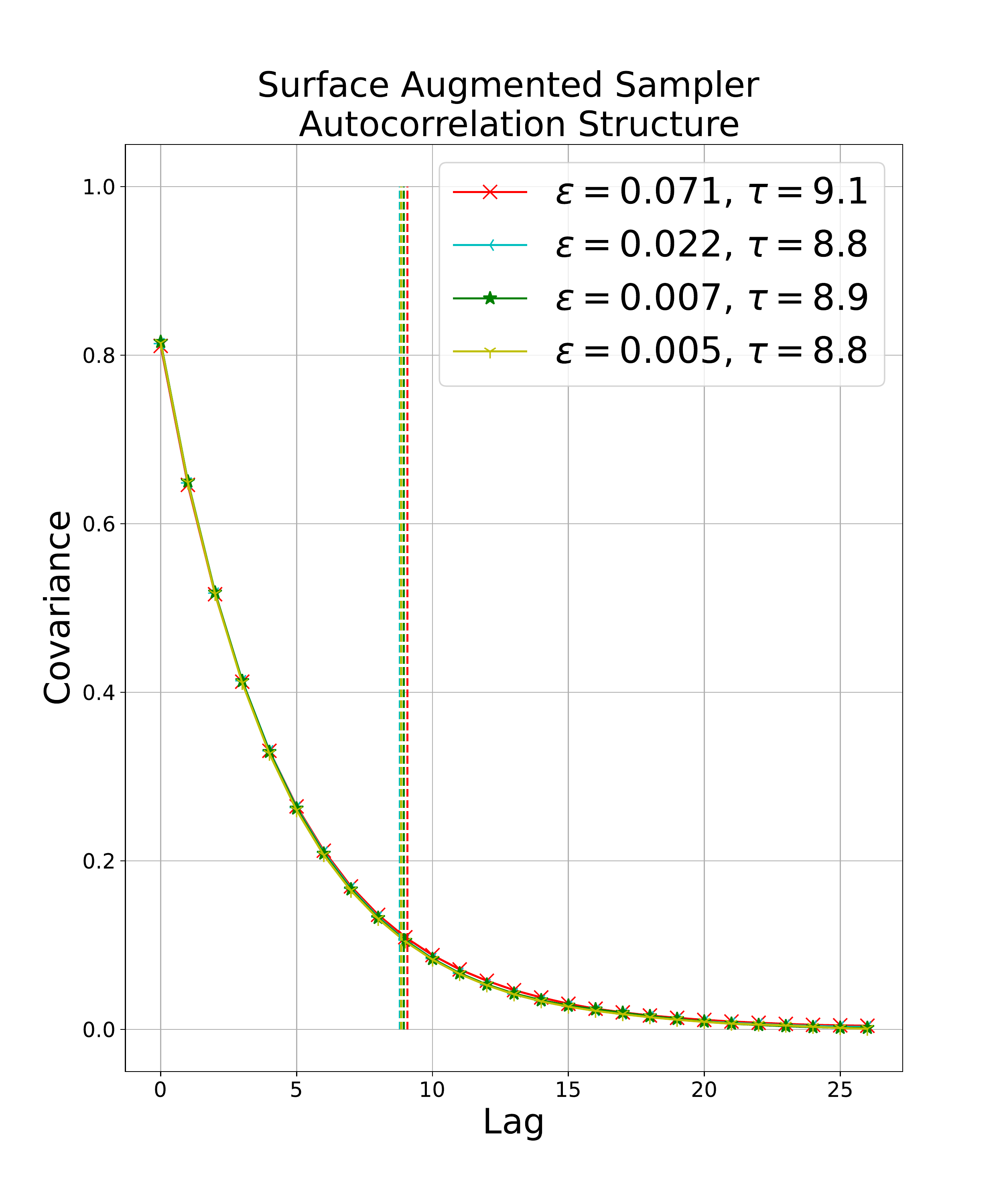}
    \caption{Ellipsoid-Sphere Model}
  \end{subfigure}
    \begin{subfigure}[b]{0.49\textwidth}
    \includegraphics[width=\textwidth]{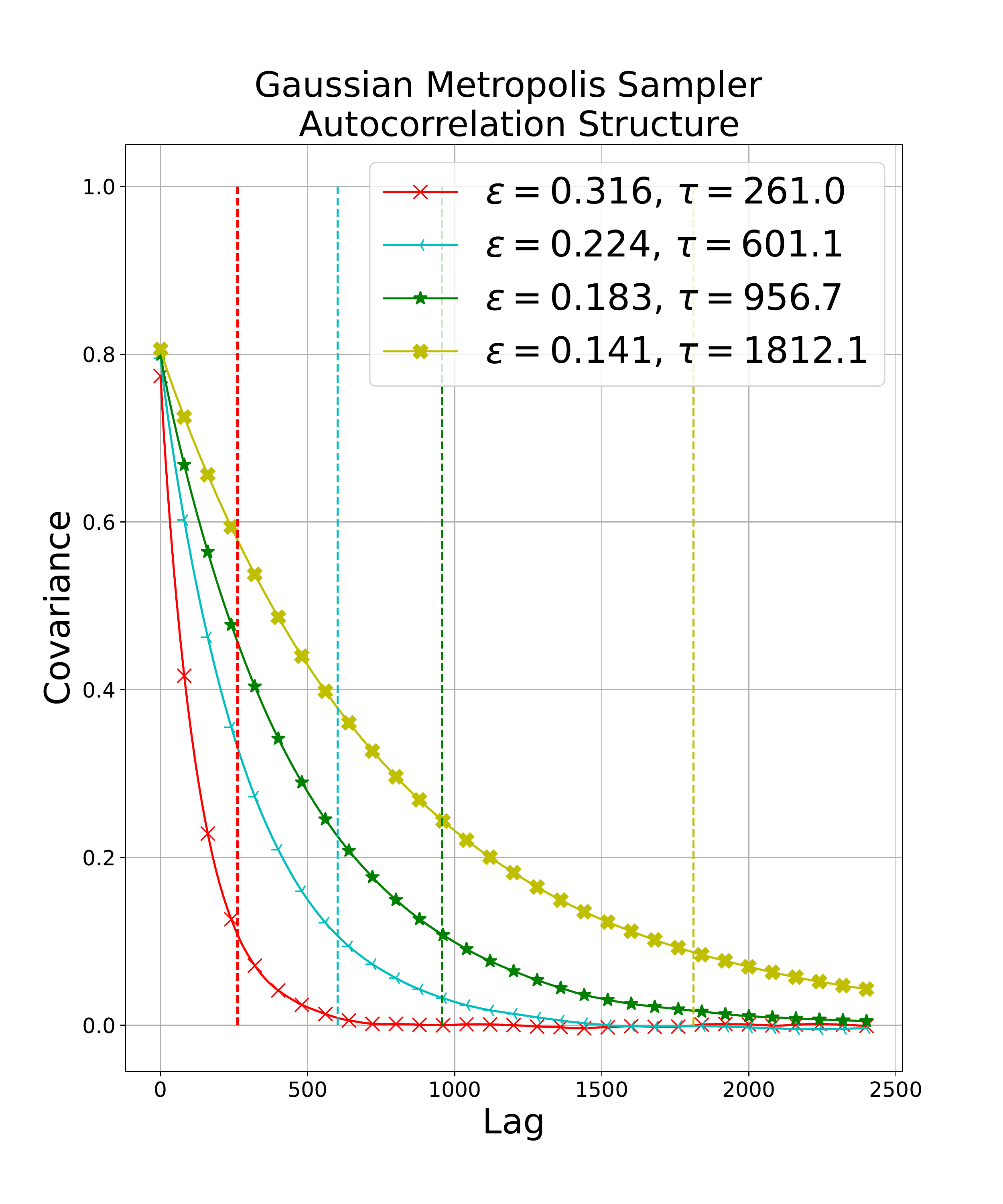}
    \caption{Ellipsoid-Sphere Model}
  \end{subfigure}
\caption{(A): Auto-correlation structure for \name\ for decreasing values of $\varepsilon$. (B): Auto-correlation structure for Isotropic Gaussian Metropolis sampler with $O(\varepsilon)$ step size for decreasing values of $\varepsilon$.} \label{fig:3}
\end{figure}

We also computed the sample auto-covariance function and auto-correlation time for decreasing values of $\varepsilon$. As before, we took inverse temperatures $\beta = 1/2\varepsilon^2 = 10^2, 10^3, 10^4, 2\times10^4$ and used $F(x) = x_1$, the projection map onto the first coordinate. See Figure \ref{fig:3} part (A) for plots of the auto-covariance function and for the auto-correlation time estimates. As expected, we found that the auto-correlation time stays approximately constant as $\varepsilon$ decreases. 

As a motivation for the \name, in Figure \ref{fig:3} part (B) we display the auto-correlation structure for the isotropic Gaussian proposal Metropolis sampler with an $O(\varepsilon)$ step size. We tested this sampler for decreasing values of $\varepsilon$. In particular, we took inverse temperatures $\beta = 1/2\varepsilon^2 = 5, 10, 15, 20$ and used $F(x) = x_1$. We set a proposal size to obtain an average acceptance probability of about $40\%$. Not surprisingly, we observe a blow-up of the auto-correlation time even for relatively small inverse temperatures (i.e., relatively large $\varepsilon$).

As a last display, in Table \ref{table:1} we show the average acceptance probability for Off and On moves for a decreasing sequence of $\varepsilon$-values. The data are generated from the \name\ using the ellipsoid-sphere model. We used step-size parameters $\sigma_\text{prp} = \sigma_\text{tan} = \sigma_\text{on} = \varepsilon$. As expected, the average acceptance probabilities converge to $1$ as $\varepsilon\downarrow 0$, confirming the analysis presented in Section \ref{sec:sec3}.

\begin{table}[ht]
\centering
\begin{tabular}{ |p{1cm}|p{2.0cm}|p{2.0cm}|  } 
\hline
Value of $\varepsilon$ & Av. Acc. Pr.\newline Off Move & Av. Acc. Pr.\newline On Move \\
\hline
0.223 & 0.768781 & 0.763992 \\
0.070 & 0.919511 & 0.919816 \\
0.022 & 0.974695 & 0.974659 \\
0.007 & 0.992233 & 0.99188  \\
0.002 & 0.997509 & 0.997627 \\
\hline
\end{tabular}
\label{Tab:1}
\vspace{0.3cm}
\caption{Average acceptance probability for Off and On moves as $\varepsilon \downarrow 0$. The chains are generated by the \name\ using the ellipsoid-sphere model. We used parameters $\sigma_\text{prp} = \sigma_\text{tan} = \sigma_\text{on} = \varepsilon$. }
\label{table:1}
\end{table}

\vspace{2cm}
\section{Verifying the Invariant Measure}\label{sec:sec4}

In this section we verify that the Markov chain generated by the \name\ preserves the target distribution $\rho$ as in \eqref{eq:2.1}, which we rewrite here for convenience of the reader
\begin{align} \label{eq:4.1}
    \rho(dx)&= \frac{1}{Z} \Big[ f_1(x) \mu_1(dx) +  f_2(x) \mu_2(dx) \Big].
\end{align}
Recall that $\mu_1$ is the reference measure for the space $(\Omega_1, \mathcal{F}_1)$ and that $\mu_2$ is the reference measure for the space $(\Omega_2, \mathcal{F}_2)$, where $\mathcal{F}_1$, $\mathcal{F}_2$ are the Borel sigma algebras for $\Omega_1$, $\Omega_2$, respectively. The target $\rho$ defines a probability distribution in ambient space $(\Omega, \mathcal{F})$, where $\Omega = \Omega_1 \cup \Omega_2$ and $\mathcal{F}$ is its Borel sigma algebra. 

Note that we can extend the measures $\mu_1$ and $\mu_2$ to the entire ambient space $(\Omega, \mathcal{F})$ so that these are mutually singular measures on $(\Omega, \mathcal{F})$ which "live" on $\Omega_1$ and $\Omega_2$, respectively. We will assume this throughout the rest of the section.

We will prove that the Markov transition kernel for the \name\ is self-adjoint in $L_\rho^2(\Omega)$. As we will see, this implies that $\rho$ is the invariant measure for the \name.  

For each pair of labels $i,j\in\{1,2\}$, we use $R_{ij}(x,dy)$ to denote the Markov transition kernel for the point-update mechanism which generates a new sample $y\in \Omega_j$, given that the current state is $x\in \Omega_i$ and that we've chosen new label $j$. This kernel can be represented as 
\begin{align}\label{eq:4.2}
    R_{ij}(x,dy) = r_{ij}(x)\delta_x(dy) + a_{ij}(x,y)h_{ij}(x,y)\mu_j(dy),
\end{align}
where $a_{ij}(x,y)$ is the Metropolis-Hastings acceptance probability \eqref{eq:2.2} and $h_{ij}(x,y)$ is the point proposal density as defined in Section \ref{sec:subsec2.2}. We also have 
\begin{align*}
\begin{split}
    r_{ij}(x) &= 1-\int_\Omega a_{ij}(x,y)h_{ij}(x,y)\mu_j(dy) \\
    &= 1-\int_{\Omega_j} a_{ij}(x,y)h_{ij}(x,y)\mu_j(dy),
\end{split}
\end{align*}
the overall probability of rejecting a proposal starting from $x$.
The kernel $R_{ij}(x,dy)$ has the following two properties: $(i)$ for each $x\in \Omega_i$, $R_{ij}(x,dy)$ defines a probability distribution on $(\Omega, \mathcal{F})$ which "lives" in $\Omega_j$ and $(ii)$ for each set $A\in \mathcal{F}$, the map $x\mapsto R_{ij}(x,A)$ is measurable on $(\Omega_i, \mathcal{F}_i)$. With this, we can write the overall transition kernel for the \name\ as
\begin{align}
\begin{split}\label{eq:4.3}
R(x,dy) = \lambda_{11}(x)&R_{11}(x,dy) + \lambda_{12}(x)R_{12}(x,dy) \\
\\
&+\lambda_{21}(x)R_{21}(x,dy) +\lambda_{22}(x)R_{22}(x,dy),
\end{split}
\end{align}
where $\lambda_{ij}(x) = \lambda_{ij} \mathds{1}_{\Omega_i}(x)$\footnote{Formally, $R_{ij}(x,dy)$ is undefined when $x\notin \Omega_i$. However, we can extend the map $x \mapsto R_{ij}(x,dy)$ to be measurable on the entire space $(\Omega, \mathcal{F})$ by setting (say) $ a_{ij}(x,y) = h_{ij}(x,y)\equiv 0$ whenever $x\notin \Omega_i$. In any case, if $x\notin \Omega_i$, then transition kernel $R_{ij}(x,dy)$ doesn't contribute to the overall kernel $R(x,dy)$ in \eqref{eq:4.3} since $R_{ij}(x,dy)$ is multiplied by $\lambda_{ij}(x) = \lambda_{ij} \mathds{1}_{\Omega_i}(x)=0$. }. Note that since $\sum_{i,j} \lambda_{ij}(x) \equiv 1$, we can conclude that $R(x,dy)$ also defines a Markov transition kernel. Equivalently, we can write
\begin{equation*}
R(x,dy) =
    \begin{cases}
        \lambda_{11} R_{11}(x,dy) + \lambda_{12} R_{12}(x,dy), & \text{if } x \in \Omega_1 \\
        \\
        \lambda_{21} R_{21}(x,dy) +\lambda_{22} R_{22}(x,dy), & \text{if } x \in \Omega_2.
    \end{cases}
\end{equation*}

We are now ready to show that $R$ is self-adjoint in $L_\rho^2(\Omega)$. First, note that the Metropolis-Hastings acceptance rule \eqref{eq:2.2} implies the detailed balance condition
\begin{align} \label{eq:4.4}
    f_i(x)\lambda_{ij}(x)h_{ij}(x,y)a_{ij}(x,y) = f_j(y)\lambda_{ji}(y)h_{ji}(y,x)a_{ji}(y,x)
\end{align}
for all pairs of labels $i,j \in \{1, 2\}$ and for any $x \in \Omega_i$ and $y \in \Omega_j$. We will use this to show that for any $f,g \in C_b(\Omega)$ we have 
\begin{align*}
    (f,Rg)_{\rho} &= \sum_{i,j} (f, R_{ij}^\lambda g)_{\rho} \\
    &= \sum_{i,j} (R_{ij}^\lambda f,g)_{\rho} = (Rf,g)_{\rho},
\end{align*}
where $R_{ij}^\lambda$ is notation for the kernel
\begin{align*}
    R_{ij}^\lambda(x,dy) = \lambda_{ij}(x)R_{ij}(x,dy),
\end{align*}
and 
\begin{align*}
    R_{ij}^\lambda f(x) = \lambda_{ij}(x) \int_{\Omega} R_{ij}(x,dy)f(y).
\end{align*}
Let $f,g \in C_b(\Omega)$. In what follows, all the integrals are intended to be over $\Omega$, so we will omit the domain subscripts for ease of notation. For any $i,j\in\{1,2\}$, we have
\begin{align}
\begin{split} \label{eq:4.5}
(f, R_{ij}^\lambda g)_{\rho} &=\iint f(x)g(y)\lambda_{ij}(x)R_{ij}(x,dy) \rho(dx) \\
\\
&=\iint f(x)g(y)\lambda_{ij}(x)\big[r_{ij}(x)\delta_x(dy) + h_{ij}(x,y)a_{ij}(x,y) \mu_j(dy) \big]\rho(dx) \\
\\
&=\int f(x)g(x)\lambda_{ij}(x)r_{ij}(x)\rho(dx) \ + \\
& \quad \quad \quad \quad \quad  \frac{1}{Z} \iint f(x)g(y)f_i(x)\lambda_{ij}(x) h_{ij}(x,y)a_{ij}(x,y)\mu_j(dy)\mu_i(dx) \\
\\
& =\int f(x)g(x)\lambda_{ij}(x)r_{ij}(x)\rho(dx) \ + \\
& \quad \quad \quad \quad \quad  \frac{1}{Z} \iint f(x)g(y)f_j(y)\lambda_{ji}(y) h_{ji}(y,x)a_{ji}(y,x)\mu_j(dy)\mu_i(dx) \\
\\
&=\int f(x)g(x)\lambda_{ij}(x)r_{ij}(x)\rho(dx) \ + \\
& \quad \quad \quad \quad \quad \frac{1}{Z} \iint f(y)g(x)f_j(x)\lambda_{ji}(x) h_{ji}(x,y)a_{ji}(x,y)\mu_j(dx)\mu_i(dy), 
\end{split}
\end{align}
where for the third equality we used that $\lambda_{ij}(x) = 0$ when $x\notin\Omega_i$ (so that only the $f_i(x)\mu_i(dx)$ part of the target $\rho(dx)$ matters in the integral), for the fourth equality we used the detailed balance condition \eqref{eq:4.4}, and for the fifth equality we interchanged the variables $x$ and $y$ in the double integral.
When $i=j = 1$, we see that
\begin{align}
\begin{split} \label{eq:4.6}
(f,R_{11}^\lambda g)_{\rho} &= \iint f(x)g(y)\lambda_{11}(x)R_{11}(x,dy) \rho(dx) \\
\\
&=\int f(x)g(x)\lambda_{11}(x)r_{11}(x)\rho(dx) \ + \\
& \quad \quad \quad \quad \quad \frac{1}{Z} \iint f(y)g(x)f_i(x)\lambda_{11}(x)h_{11}(x,y)a_{11}(x,y)\mu_i(dx)\mu_i(dy) \\
\\
&= \iint f(y)g(x)\lambda_{11}(x)R_{11}(x,dy) \rho(dx) = (R_{11}^\lambda f,g)_{\rho},
\end{split}
\end{align}
where for the second equality we used \eqref{eq:4.5}. Therefore, $R_{11}^\lambda(x,dy)$ is self-adjoint. The case when $i=j=2$ is similar.

We are left to verify that 
\begin{align} \label{eq:4.7}
(f,R_{12}^\lambda g)_{\rho} +(f,R_{21}^\lambda g)_{\rho} =(R_{12}^\lambda f,g)_{\rho} +(R_{21}^\lambda f,g)_{\rho}.
\end{align}
Note that by \eqref{eq:4.5},  we can write
\begin{align}
\begin{split} \label{eq:4.8}
(f,R_{12}^\lambda g)_{\rho} &=\int f(x)g(x)\lambda_{12}(x)r_{12}(x)\rho(dx) \ + \\
&\ \ \ \ \ \ \ \ \ \ \ \ \frac{1}{Z}\iint f(y)g(x)f_2(x)\lambda_{21}(x)h_{21}(x,y)a_{21}(x,y)\mu_2(dx)\mu_1(dy),
\end{split}
\end{align}
and
\begin{align}
\begin{split} \label{eq:4.9}
(f,R_{21}^\lambda g)_{\rho} &=\int f(x)g(x)\lambda_{21}(x)r_{21}(x)\rho(dx) \ + \\
&\ \ \ \ \ \ \ \ \ \ \ \ \frac{1}{Z} \iint f(y)g(x)f_1(x)\lambda_{12}(x)h_{12}(x,y)a_{12}(x,y)\mu_1(dx)\mu_2(dy).
\end{split}
\end{align}
Finally, by simply applying the definitions we get
\begin{align}
\begin{split} \label{eq:4.10}
(R_{12}^\lambda f,g)_{\rho} &=\iint  g(x)f(y)\lambda_{12}(x)R_{12}(x,dy) \rho(dx) \\
\\
&=\int g(x)f(x)\lambda_{12}(x)r_{12}(x)\rho(dx) \ + \\
&\ \ \ \ \ \ \ \ \ \ \ \ \frac{1}{Z}  \iint f(y)g(x)f_1(x)\lambda_{12}(x)h_{12}(x,y)a_{12}(x,y)\mu_2(dy)\mu_1(dx), 
\end{split}
\end{align}
and
\begin{align}
\begin{split} \label{eq:4.11}
(R_{21}^\lambda f,g)_{\rho} &=\iint  g(x)f(y)\lambda_{21}(x)R_{21}(x,dy) \rho(dx) \ + \\
\\
&=\int g(x)f(x)\lambda_{21}(x)r_{21}(x)\rho(dx) \\
&\ \ \ \ \ \ \ \ \ \ \ \ \frac{1}{Z} \iint f(y)g(x)f_2(x)\lambda_{21}(x)h_{21}(x,y)a_{21}(x,y)\mu_1(dy)\mu_2(dx),
\end{split}
\end{align}
By combining \eqref{eq:4.8}, \eqref{eq:4.9}, \eqref{eq:4.10} and \eqref{eq:4.11}, we see that \eqref{eq:4.7} holds. We thus proved that $(f, Rg)_\rho = (Rf, g)_\rho$ for any $f, g\in C_b(\Omega)$. Since $C_b(\Omega)$ is dense in $L^2_\rho(\Omega)$, we conclude that $R$ is self-adjoint in $L^2_\rho(\Omega)$.

As a last step, we show that self-adjointness of $R$ in $L^2_\rho(\Omega)$ implies that $R$ preserves the target distribution $\rho$ as in \eqref{eq:4.1}. Indeed, by self-adjointness of $R$ we have
\begin{align} \label{eq:4.12}
    (f, 1)_\rho = (f, R1)_\rho = (Rf, 1)_\rho , 
\end{align} 
for any $f \in L^2_\rho(\Omega)$. In particular, if we take any measurable set $A$ and set $f = \mathds{1}_{A}$ in \eqref{eq:4.12}, we get
\begin{align} \label{eq:4.13}
   \rho(A) = \int R(x,A) \rho(dx).
\end{align}

\appendix

\section{Target Distribution in the Soft Constraints Limit}\label{sec:A}

We will give a semi-rigorous argument for \eqref{eq:1.4}, \eqref{eq:1.5} that uses the \textit{co-area} formula. Take any function $\varphi \in C_c^\infty(\mathbb{R}^{d_a})$. Use $S_a$ to denote the set $S_a = \big\{ x\in\mathbb{R}^{d_a} : q(x) = a \big\}$ and $\sigma_a(dx)$ to denote the $d$-dimensional Hausdorff measure on $S_a$. Let $R(x) = \nabla q(x)^T \nabla q(x) $, and define the functions $g(a) = \int_{S_a} \varphi(x) |R(x)|^{-1/2} \sigma_a(dx)$ and $h(a) = \int_{S_a} |R(x)|^{-1/2} \sigma_a(dx)$. First, we observe that
\begin{align}
\begin{split} \label{eq:A1}
    Z_\varepsilon = \int_{\mathbb{R}^{d_a}} &e^{\frac{U(x)}{2\varepsilon^2}} dx = \int_{\mathbb{R}^m} \int_{S_a} e^{\frac{U(x)}{2\varepsilon^2}} |R(x)|^{-1/2} \sigma_a(dx) da \\
    \\
    &= \int_{\mathbb{R}^m}  e^{-\frac{|a|^2}{2\varepsilon^2}} h(a) da \ \sim \ h(0)(2 \pi)^{m/2}\varepsilon^m, \quad \text{as} \ \varepsilon \downarrow 0,
\end{split}
\end{align}
where the second equality is just the co-area formula, in the third equality we used that $U(x) = |a|^2$ when $x\in S_a$, and the last step follows from Laplace's method. Then, 
\begin{align}
\begin{split} \label{eq:A2}
    \int_{\mathbb{R}^{d_a}} \varphi(x) \pi_\varepsilon(dx) &= \int_{\mathbb{R}^m} \int_{S_a} \varphi(x) \frac{1}{Z_\varepsilon}e^{-\frac{U(x)}{2\varepsilon^2}} |R(x)|^{-1/2} \sigma_a(dx) da \\
    \\
    &= \frac{1}{Z_\varepsilon} \int_{\mathbb{R}^m} e^{-\frac{|a|^2}{2\varepsilon^2}} g(a) da \\
    \\
    &\approx \frac{1}{h(0)(2 \pi)^{m/2}\varepsilon^m} \int_{\mathbb{R}^m} e^{-\frac{|a|^2}{2\varepsilon^2}} g(a) da \\
    \\
    &\xrightarrow[]{\varepsilon \downarrow 0} \ \frac{1}{h(0)} g(0) \\
    \\
    &= \frac{1}{h(0)} \int_{S} \varphi(x) |R(x)|^{-1/2} \sigma(dx) \\
    \\
    &= \int_{S} \varphi(x) \pi_s(dx),
\end{split}
\end{align}
where $\pi_s(x) = |R(x)|^{-1/2} \sigma(dx) / h(0)$. The first equality in \eqref{eq:A2} is (once again) the co-area formula. In the third line of \eqref{eq:A2} we used the asymptotic approximation \eqref{eq:A1}. The limit in the fourth line holds by standard approximate identity arguments. 

\nocite{H19}
\AtNextBibliography{\scriptsize}
\printbibliography

\end{document}